\documentclass[12pt,a4paper]{article}
\usepackage[]{amsmath}
\usepackage[]{amssymb}
\usepackage[]{hyperref}
\usepackage[]{graphicx}
\usepackage{epstopdf}
\newcommand{\sg}{\mathfrak{s}}
\newcommand{\cg}{\mathfrak{c}}

\begin{document}
\begin{flushright}
INR-TH-2016-022
\end{flushright}

\vspace{10pt}
\begin{center}
  {\LARGE \bf Perturbations on and off de~Sitter
brane\\ [0.3cm] in anti-de~Sitter
bulk} \\
\vspace{20pt}
M.~Libanov$^{a,b}$ and V.~Rubakov$^{a,c}$ \\
\vspace{15pt}

$^a$\textit{
Institute for Nuclear Research of
         the Russian Academy of Sciences,\\  60th October Anniversary
  Prospect, 7a, 117312 Moscow, Russia}\\
\vspace{5pt}

$^b$\textit{Moscow Institute of Physics and Technology,\\
Institutskii per., 9, 141700, Dolgoprudny, Moscow Region, Russia
}

\vspace{5pt}

$^c$\textit{Department of Particle Physics and Cosmology,
Physics Faculty, M.V.~Moscow State University\\ Vorobjevy Gory,
119991, Moscow, Russia}

    \end{center}
    \vspace{5pt}

\begin{abstract}
Motivated by holographic models of (pseudo)conformal Universe,
we carry out complete analysis of linearized metric perturbations in
the time-dependent two-brane
setup of the Lykken-Randall type. We present the
equations of motion for the scalar, vector and tensor perturbations and
identify light modes in the spectrum, which are scalar radion and
transverse-traceless graviton. We show that there are no other modes in
the discrete part of the spectrum.
We pay special attention to properties of
light modes and show, in particular, that the radion has red power
spectrum at late times, as anticipated on holographic grounds.
Unlike the graviton, the radion survives in the single-brane limit,
when one of the branes is sent to the adS boundary.
These properties imply that  potentially observable features characteristic of
the 4d (pseudo)conformal cosmology, such as  statistical anisotropy
and specific shapes of
non-Gaussianity, are inherent also in  holographic conformal
models
as well as in  brane world
inflation.
\end{abstract}

\section{Introduction}

Some time ago it has been pointed out that conformal symmetry $SO(4,2)$
broken down to de Sitter $SO(4,1)$ in the early Universe may be
responsible for the generation of the (nearly) flat spectrum of scalar
cosmological perturbations~\cite{Rubakov:2009np, Creminelli:2010ba,
Hinterbichler:2011qk, Hinterbichler:2012fr} (see
Ref.~\cite{Libanov:2015iwa} for a review). The main ingredient of the
(pseudo)conformal scenarios  is the expectation value of a scalar operator
$\mathcal{ O}$ of non-zero conformal weight $\triangle$ which depends on
time $\tau$ and gives rise to symmetry breaking,
\begin{equation}
\langle \mathcal{ O}\rangle \propto \frac{1}{(-\tau )^{\triangle}}\,,
\label{Eq/Pg1/1:dr}
\end{equation}
where $\tau <0$. It is assumed also that: (i) space-time is effectively
Minkowskian during the rolling stage (\ref{Eq/Pg1/1:dr}); (ii) there is
another scalar field of zero effective conformal weight in this
background, whose perturbations automatically have flat power
spectrum\footnote{Weak explicit breaking of conformal invariance yields
small tilt in this spectrum~\cite{Osipov:2010ee}.}; (iii) the
perturbations of the latter field are converted into the adiabatic scalar
perturbations at some later stage.

A peculiarity inherent in the (pseudo)conformal mechanism is that the
perturbations of $\mathcal{ O}$ have red power spectrum,
\begin{equation}
\mathcal{ P}_{\delta \mathcal{ O}} \propto p^{-2}\;.
\label{Eq/Pg1/2:dr}
\end{equation}
This feature leads to potentially observable predictions, such as specific
shapes of non-Gaussianity~\cite{Libanov:2011hh, Creminelli:2012qr,
nongauss} and statistical anisotropy~\cite{Libanov:2011hh,
Creminelli:2012qr, anisotropy, constraniso}. It is worth emphasizing that
many of these properties are direct consequences of the symmetry breaking
pattern $SO(4,2)\to SO(4,1)$~\cite{Creminelli:2012qr,
Hinterbichler:2012mv}.

Further development of the (pseudo)conformal scenario involves holography.
It has been pointed out that conformal rolling (\ref{Eq/Pg1/1:dr}) in the
boundary theory is dual to motion of a domain wall in the adS$_5$
background~\cite{Hinterbichler:2014tka, Libanov:2014nla}. This motion
corresponds to spatially homogeneous transition from a false vacuum to a
true one. One generalizes this construction further and considers
nucleation and subsequent growth, in adS$_{5}$, of a bubble of the true
scalar field vacuum surrounded by the false vacuum. From the viewpoint of
the boundary CFT, this process corresponds to the (spatially
inhomogeneous) Fubini--Lipatov tunneling transition and subsequent
real-time development of an instability of a conformally invariant
vacuum~\cite{Libanov:2015mha}. In the holographic approach the position of
the moving domain wall plays the role of the operator $\mathcal{ O}$ whose
perturbations again have  red power spectrum (\ref{Eq/Pg1/2:dr}).

It is worth noting that the analysis of perturbations in these holographic
constructions has not included so far the effects of dynamical 5d gravity:
the back reaction of the domain wall perturbations on the background
 adS$_5$ has been neglected. Clearly, it is of interest to understand
whether or not the power spectrum (\ref{Eq/Pg1/2:dr}) gets modified by the
  effects of dynamical 5d gravity; this is one of the issues we address in
this paper (within the thin brane approximation).

In fact, various brane-gravity systems in adS$_5$ background have been
studied in the context of brane-world models with large and infinite extra
dimensions (for a review see, e.g., Ref.~\cite{Rubakov:2001kp}). In
particular, the linearized metric perturbations have been analyzed in the
framework of the static Randall-Sundrum I (RS1) model with
$S^{1}/\mathbb{Z}_{2}$ orbifold extra  dimension and two 3-branes (one
with positive and another with negative tension) residing at its
boundaries~\cite{Randall:1999ee}. It has been
shown~\cite{Charmousis:1999rg}  that apart from massless four-dimensional
graviton (whose wave function is peaked at the positive tension brane) and
the corresponding Kaluza-Klein tower,  the perturbations  contain a
massless four-dimensional scalar field, radion, which corresponds to the
relative motion of the branes. The radion wave function is peaked at the
negative tension brane.  In Ref.~\cite{PRZ} the metric perturbations have
been studied in a more general static setup~\cite{Lykken:1999nb} where the
assumption of the $\mathbb{Z}_{2}$ symmetry across the visible brane has
been dropped. It has been shown that the radion becomes a ghost in some
region of the parameter space which, in particular, includes the setup of
Refs.~\cite{Charmousis:1999rg, Gregory:2000jc} where graviton  is
quasi-localized due to the warped geometry of the bulk. Similar results
were obtained in Ref.~\cite{Dubovsky:2003pn} where  effects of the induced
Einstein term on the brane(s) have been considered.

It is worth recalling that the static brane world setups are possible only
if certain fine tuning relation(s)  between the bulk cosmological
constant(s)  and the brane tension(s) are satisfied. If these conditions
are not met, the background in general depends on time. In the simple
one-brane setup in a frame where an observer is at rest with respect to
the bulk, the bulk geometry is (locally) static and anti de Sitter while
the brane  moves along the extra dimension, and the brane induced metric
corresponds to de Sitter space~\cite{Kraus:1999it, Ida:1999ui,
Cvetic:1999ec, Mukohyama:1999wi, Bowcock:2000cq, Gorbunov:2001ge}. On the
other hand, from the viewpoint of an observer located on the brane, the
induced geometry of the brane is still de Sitter, while the bulk metric
becomes time-dependent.

The above discussion suggests that in the dynamical background, the radion
(which is a massless scalar field in the case of the static background)
becomes a scalar field with red power spectrum (\ref{Eq/Pg1/2:dr}). The
radion in the RS1 setup with a slice of adS$_5$ bound by two dS$_4$ branes
(one with positive tension and another with negative tension) was studied
in Refs.~\cite{Gen:2000nu, Binetruy:2001tc, Chacko:2001em, Gen:2002rb}
(see also Ref.~\cite{Chiba:2000rr}) with the result that the radion
perturbations have red power spectrum indeed.

In this paper we consider the linearized metric perturbations in a more
general spatially homogeneous thin-brane setup of the Lykken-Randall
type~\cite{Lykken:1999nb} with the relaxed fine tuning conditions, and
hence with time-dependent background. Although we consider for
completeness the case when one of the branes has negative tension, our
primary interest is the model with both branes having positive tensions.
This setup is more reminiscent of the holographic description of the
conformal vacuum decay, albeit it is spatially homogeneous and does not
involve a scalar field in the bulk. We will pay  special attention to the
radion and show that its equation of motion indeed leads to  red power
spectrum which has precisely the form~(\ref{Eq/Pg1/2:dr}). Importantly,
there are no other scalar modes bound to any of the branes: all other
modes belong to continuous spectrum. Similar situation occurs in the
tensor sector, which contains one mode bound to the UV brane (which is
essentially the Randall--Sundrum graviton) and modes from continuum. One
of our main purposes is to see what happens in a model with a {\it single}
brane, that generalizes the model of Ref.~\cite{Libanov:2014nla} in the
sense that it includes effects of the 5d gravity. In this context, the
Lykken--Randall UV brane is viewed as a regularization tool, so we send it
to the adS$_5$ boundary in the end. We find that the radion perturbations
do not decouple in this limit and still have the power
spectrum~(\ref{Eq/Pg1/2:dr}). Thus, the potentially observable features
of the (pseudo)conformal universe~\cite{Libanov:2015iwa} hold for the de
Sitter brane moving in the 5d bulk.

This paper is organized as follows. In
Sec.~\ref{Section/Pg1/1:dyn_radion/The Setup} we describe the two-brane
setup. In Sec.~\ref{Section/Pg4/1:dyn_radion/Perturbations} we consider
general metric perturbations and fix the gauge. We also identify a radion
mode which corresponds to relative brane fluctuation. In
Secs.~\ref{Section/Pg8/1:dyn_radion/Einstein equations} and
\ref{Section/Pg12/1:dyn_radion/Linearized Israel conditions} we present
the linearized Einstein equations and Israel junction conditions. In
Sec.~\ref{Section/Pg11/1:dyn_radion_aDs4/The Einstein equations solution}
we solve the full set of equations in scalar, vector and tensor sectors of
the metric perturbations. In
Sec.~\ref{Section/Pg19/1:dyn_radion_adS4_2_branes/Light modes effective
action} we construct effective actions for light modes, radion and
graviton. We discuss the properties of the radion and show that its
perturbations have red power spectrum. We consider the single brane limit
and show that the radion does not decouple and that the spectrum of its
perturbations  remains red. We conclude in
Sec.~\ref{Section/Pg32/1:dr/Conclusion}.

\section{Setup and background}
\label{Section/Pg1/1:dyn_radion/The Setup}

\begin{figure}[t]
\begin{center}
\includegraphics[ width=175.34mm, height=79.03mm,
]{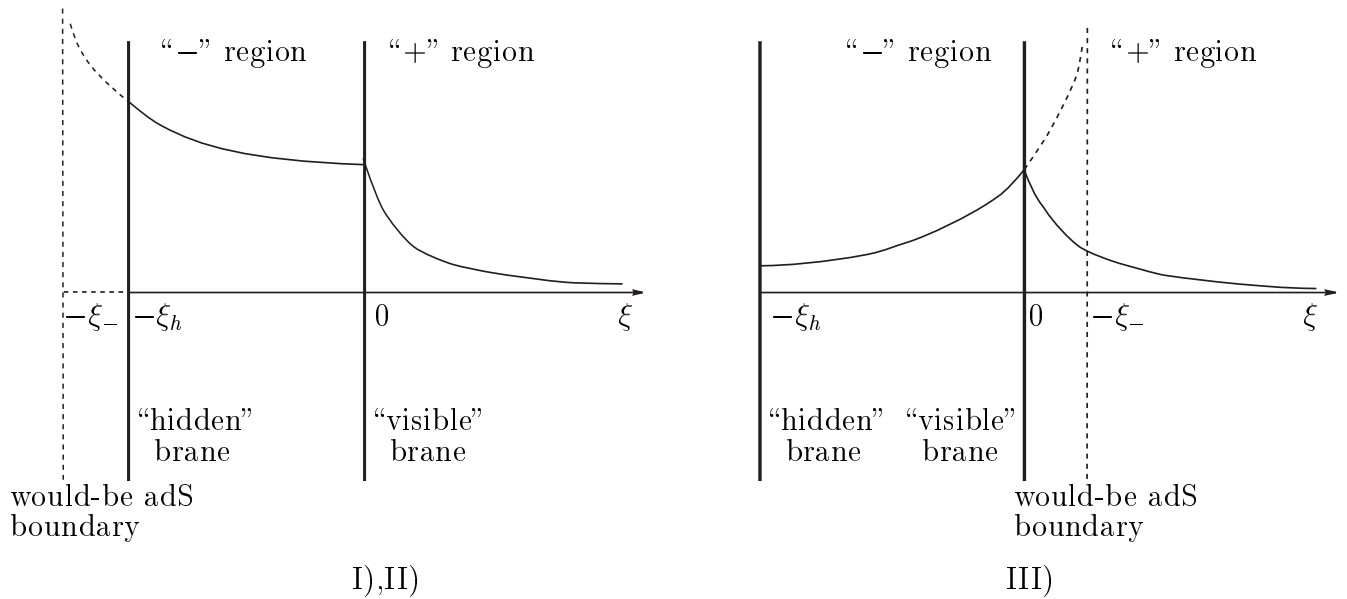}
\end{center}
\caption{Two-brane setup. Solid line shows the warp factor $(k\sg)^{-1}$.
Left and right panels correspond to the cases I), II) ($k_{\pm}>0$) and
III) ($k_{+}>0\,\ k_{-}<0$), see the text after eq. (\ref{Eq/Pg4/1:dr}).
\label{Fig/Pg7/1:ml-vr-v3}
}
\end{figure}

We consider the $(d+2)$-dimensional background with the metric
\begin{equation}
ds^{2}=\frac{1}{k_{\pm}^{2}\sg^{2}_{\pm} }\left(\frac{\eta _{\mu \nu
}}{\tau ^{2}}dx^{\mu }d{x^{\nu }}-d\xi ^{2} \right)\,,
\label{Eq/Pg1/1:dyn_radion}
\end{equation}
where $k_{\pm}$ and $\xi _{\pm}$ are constants, $\tau \equiv x^{0} <0$,
and hereafter we use the notations
\[
\sg_{\pm}=\sinh(\xi +\xi _{\pm})\,,\ \ \ \ \cg_{\pm}=\cosh(\xi +\xi
_{\pm})\,.
\]
This metric is a solution of the $(d+2)$-dimensional gravity with two
thin $d$-brane sources,
\begin{equation}
S=-M^{d}\int \limits_{}^{} d^{d+2}X\sqrt{|g|}R-\Lambda \int
\limits_{}^{}d^{d+2}X\sqrt{|g|} - \sum \limits_{i=1}^{2}\lambda_{i} \int
\limits_{}^{}d^{d+2}X\sqrt{|\gamma^{(i)}|}\delta (\xi-\xi _{i} ),
\label{Eq/Pg1/2:dyn_radion}
\end{equation}
where $g_{AB}$ is the bulk metric and $\gamma _{\mu \nu }^{(i)}$ is the
metric induced on the $i$-th brane. The (negative) $(d+2)$-dimensional
cosmological constants  may be different in different domains of the bulk
space, separated by $d$-branes. We consider the  model of the
Lykken-Randall type with two branes~\cite{Lykken:1999nb}. The first
(``hidden'', or UV) brane is placed at the fixed point $\xi =-\xi _{h}$ of
the $\mathbb{Z}_{2}$ orbifold symmetry $\xi \to -2\xi _{h}-\xi $. The
second (``visible'') brane separating two domains with different $\Lambda
_{\pm}$ is at $\xi =\xi_v =0$. We will relate the parameters $k_{\pm}$,
$\xi_{\pm}$ of the solution to the parameters of the action in due course.
In what follows the domain between two branes $-\xi _{h}<\xi <0$ is
referred to as ``$-$'' region while the domain $\xi >0$ is ``$+$'' region,
see Fig.~\ref{Fig/Pg7/1:ml-vr-v3}.

As a side remark, it is instructive to consider the reference frame in
which the bulk geometry is (locally) static. The
coordinates in this frame, $t$ and $r$, are related to $\tau $ and $\xi $
as follows:
\begin{eqnarray}
t&=&\tau \frac{\cg_{\pm}}{\mathrm{c}_{\pm}} < 0\,,\nonumber\\
r&=&-\tau k_{\pm} \sg_{\pm}\,. \nonumber
\end{eqnarray}
Hereafter
\[
\mathrm{s}_{\pm}=\sinh\xi _{\pm}\,,\ \ \ \ \mathrm{c}_{\pm}=\cosh\xi
_{\pm}\, .
\]
Due to one of the Israel junction conditions, one has
$k_{-}\mathrm{s_{-}}=k_{+}\mathrm{s}_{+} \equiv H$ (see below), where $H$
is the Hubble parameter on the visible brane given
by~(\ref{Eq/Pg10/1:dyn_radion_aDs4}). So, the coordinates $t$, $r$ are
continuous across the visible brane. In these coordinates the bulk
geometry is described by
\begin{equation}
ds^{2}=\frac{1}{k_{\pm}^{2}r^{2}}(\mathrm{c}_{\pm}^{2}k_{\pm}^{2}dt^{2}-
k_{\pm}^{2}d\mathbf{x}^{2}-dr^{2})\,,
\label{Eq/Pg1/2:sl2st}
\end{equation}
which is the Poincar\'e metric in the two patches of adS$_{d+2}$ with
different cosmological constants. In this frame the branes are moving.
Their positions are given by
\[
r_{v}(t)=-Ht\,,\ \ \ \
r_{h}(t)=-Ht\frac{\mathrm{c_{-}}}{\mathrm{s_{-}}}\cdot
\frac{\mathrm{s_{h}}}{\mathrm{c}_{h}}\, ,
\]
where
\[
\mathrm{s}_{h}=\sinh(\xi _{-}-\xi _{h})\,,\ \ \ \
\mathrm{c}_{h}=\cosh(\xi _{-}-\xi _{h})\; .
\]
Thus, our setup is similar to Refs.~\cite{Hinterbichler:2014tka,
Libanov:2014nla}, where the domain wall moves along $r/t = \mbox{const}$
in the Poincar\'e coordinates. We do not use the coordinates $r$, $t$ in
what follows.

The components of the unperturbed Ricci tensor, calculated with the metric
(\ref{Eq/Pg1/1:dyn_radion}), are (hereafter we skip (sub-)superscript
``$\pm$'' where this does not lead to an ambiguity),
\[
R_{\mu \nu }=\frac{(d+1)}{\tau ^{2}\sg^{2}}\eta _{\mu \nu }\,,\ \ \ \
R_{\mu \xi }=0\,, \ \ \ \ R_{\xi \xi }=-\frac{d+1}{\sg^{2}}\,,
\]
and satisfy the Einstein equations in the bulk
\begin{equation}
R_{AB}=-\frac{\Lambda _{\pm}}{2M^{d}}\cdot\frac{2}{d}\cdot g_{AB},
\label{Eq/Pg1/3:dyn_radion}
\end{equation}
provided that the values of the inverse adS radii $k_{\pm}$ are related to
the cosmological constants,
\begin{equation}
|k_{\pm}|=\sqrt{-\frac{\Lambda _{\pm}}{M^{d}d(d+1)}}.
\label{Eq/Pg1/4:dyn_radion}
\end{equation}
We take $k_{+}$ positive (without loss of generality) and assume that
there is no boundary at $\xi >0$, which implies
\begin{equation}
\xi_+ >0 \; , \;\;\;\; k_+ > 0 \; .
\label{jul7-16-1}
\end{equation}
Other sign conventions are that in the case $k_{-}>0$ the hidden brane
screens the adS boundary at $\xi \to -\xi _{-} $, while for $k_{-}<0$ one
can push the hidden brane to infinity. Hence, the two options are
\begin{subequations}
\label{jul7-16-2}
\begin{align}
&\xi_- >0 \; , \;\;\;\; k_- > 0 \; \;\;\; \mbox{(and}\;\; \xi_h < \xi_-)
\; ,
\label{jul7-16-2a}\\
& \xi_- < 0 \; , \;\;\;\; k_- < 0 \; .
\label{jul7-16-2b}
\end{align}
\end{subequations}

To proceed, we make use of the Israel junction
 conditions~\cite{Israel:1966rt} to determine the boundary conditions on
the branes. It is worth recalling the definition of the extrinsic
curvature $K_{\mu \nu }$ and the induced metric on the branes (see, e.g.,
Ref.~\cite{Berezin:1987bc}). Let $F(X^{A})=0$ be the equation of time-like
hypersurface $\Sigma $ and $y^{\mu }$  be coordinates on it. We introduce
tangent vectors to $\Sigma $
\[
e^{A}_{\mu }=\frac{\partial X^{A}}{\partial y^{\mu }}\Bigg|_{\Sigma }
\,,
\]
and the normal unit outer vector
\[
n_{A}=-\frac{\partial _{A}F}{\sqrt{|\partial _{B}F\partial ^{B}F}|}\,,
\]
which is space-like. Then the induced metric and the extrinsic
curvature are given by
\begin{eqnarray}
\gamma _{\mu \nu }&=&e_{\mu }^{A}e_{\nu
}^{B}g_{AB}\Big|_{\Sigma}\,,\nonumber\\
K_{\mu \nu }&=&e^{A}_{\mu }e^{B}_{\nu }D_{A}n_{B}\Big|_{\Sigma }\,.
\nonumber
\end{eqnarray}

The Israel junction conditions at each of the branes are
\begin{eqnarray}
\triangle \gamma _{\mu \nu }^{(i)}&=&0\,, \nonumber \\
\triangle K_{\mu \nu }^{(i)}&=&-\frac{\lambda ^{(i)}}{2dM^{d}} \gamma
_{\mu \nu }^{(i)}\, . \nonumber
\end{eqnarray}
Hereafter $\triangle$ denotes a jump of the corresponding quantity across
the brane from $\xi >\xi ^{(i)}$ to $\xi <\xi ^{(i)}$.

Due to $\mathbb{Z}_{2}$ symmetry, the continuity of the induced metric
\[
\gamma _{\mu \nu }^{(h)}=\frac{1}{k_{-}^{2}\mathrm{s}_{h}^{2}\tau
^{2}}\eta _{\mu \nu }\,,
\]
on the hidden brane $(\xi =-\xi _{h})$ is trivially satisfied. The jump
of the extrinsic curvature is given by
\begin{equation}
\triangle \gamma ^{\nu \rho }_{(h)}K_{\rho \mu
}^{(h)}=-2k_{-}\mathrm{c}_{h}\delta ^{\nu }_{\mu }=-\frac{\lambda
^{(h)}}{2dM^{d}}\delta _{\mu }^{\nu }\,.
\label{Eq/Pg3/1:dr}
\end{equation}
Eq. (\ref{Eq/Pg3/1:dr}) yields a relation between the hidden brane tension
and its position:
\begin{equation}
\lambda ^{(h)} =4dM^{d}k_{-}\mathrm{c}_{h}\,.
\label{Eq/Pg3/2:dr}
\end{equation}

On the visible brane the induced metric and the extrinsic curvature are
\begin{eqnarray}
\gamma _{\mu \nu }^{(v),\pm}&=&\frac{1}{k^{2}_{\pm}\tau
^{2}\mathrm{s_{\pm}}^{2}}\eta_{\mu \nu} \,,
\label{Eq/Pg3/3:dr}\\
\gamma ^{\nu \rho }_{(v),\pm}K_{\rho \mu }^{(v),\pm}
&=&-k_{\pm}\mathrm{c_{\pm}}\delta _{\mu }^{\nu } \, .
\label{Eq/Pg3/4:dr}
\end{eqnarray}
Then the Israel junction conditions become (using the sign conventions
(\ref{jul7-16-1}) and (\ref{jul7-16-2}))
\begin{eqnarray}
k_{-}\mathrm{s_{-}}=k_{+}\mathrm{s}_{+}\equiv H\,,
\label{Eq/Pg2/1:dyn_radion_aDs4}\\
k_{+}\mathrm{c_{+}}-k_{-}\mathrm{c_{-}}=\sigma \,,
\label{Eq/Pg9/5:dyn_radion_aDs4}
\end{eqnarray}
where
\[
\sigma =\frac{\lambda^{(v)}}{2dM^{d}} \; .
\]
By solving these equations one gets a relation between the Hubble
constant $H$ and the parameters $k_{\pm}$ and $\sigma $,
\begin{equation}
H^{2}=k^{2}\mathrm{s}^{2} =\frac{\left( (k_{+}+k_{-})^{2}-\sigma ^{2}
\right)\left((k_{+}-k_{-})^{2}-\sigma ^{2} \right)}{4\sigma ^{2}}\,,
\label{Eq/Pg10/1:dyn_radion_aDs4}
\end{equation}
and finds
\begin{equation}
\xi _{\pm}=\mathrm{arcsinh}\frac{H}{k_{\pm}}
\,.
\label{Eq/Pg10/2:dyn_radion_aDs4}
\end{equation}
Three remarks are in order. First, the fact that the solutions
(\ref{Eq/Pg10/2:dyn_radion_aDs4}) exist confirms that the metric
(\ref{Eq/Pg1/1:dyn_radion}) with constant $\xi _{\pm}$ is a solution to
the Einstein equations and Israel junction conditions. Second, as it
follows from (\ref{Eq/Pg3/3:dr}), the brane is in the de Sitter regime
with the Hubble parameter $H\geq 0$ given by
(\ref{Eq/Pg10/1:dyn_radion_aDs4}). Our primary interest in the case $\tau
<0$ which corresponds to expanding branes. Most of our formulas, however,
are valid also for contracting branes, $\tau >0$. Third, substituting
(\ref{Eq/Pg10/1:dyn_radion_aDs4}), (\ref{Eq/Pg10/2:dyn_radion_aDs4}) into
eq.(\ref{Eq/Pg9/5:dyn_radion_aDs4}) one gets,
\begin{equation}
\mathrm{sign}(k_{+})|k_{+}^{2}-k_{-}^{2}+\sigma ^{2}| -
\mathrm{sign}(k_{-})|k_{+}^{2}-k_{-}^{2}-\sigma ^{2}|=2|\sigma |\sigma\,.
\label{Eq/Pg4/1:dr}
\end{equation}
For $k_{+}>0$ this equation together with the condition $H \geq 0$
 leads, in general, to the following three cases:
\begin{itemize}
\item[I)] $k_{+}>0$, $k_{-}> 0$, $\sigma >0$~~ $\Longrightarrow$ ~~
$k_{+}\geq k_{-}+\sigma$, ~~~ $\xi _{+},\xi _{-}>0$;
\item[II)] $k_{+}> 0$, $k_{-}>0$, $\sigma <0$~~ $\Longrightarrow$~~ $0 <
 k_{+}\leq k_{-}-|\sigma |$,  ~~~ $\xi _{+},\xi _{-}>0$
\item[III)] $k_{+}> 0$, $k_{-}< 0$, $\sigma >0$~~ $\Longrightarrow$~~ $0<
k_{+}\leq \sigma -|k_{-}|$, ~~~ $\xi _{+}>0$, $\xi _{-}<0$\; ,
 \end{itemize}
which is consistent with our sign convention (\ref{jul7-16-1}),
(\ref{jul7-16-2}).

To conclude this section let us consider the static limit $H\to 0$. To
this end we require that the resulting background metric  takes the form
(\ref{Eq/Pg1/2:sl2st}) with $\mathrm{c}_{\pm}\to 1$, and  that the visible
brane is located at $r_{v}=1$ while the hidden one is at $r_{h}$. In  all
three cases I)-III) discussed above the limit $H\to 0$ is approached in
the following regime (see (\ref{Eq/Pg9/5:dyn_radion_aDs4}),
(\ref{Eq/Pg10/2:dyn_radion_aDs4})):
\begin{equation}
\xi _{\pm}\simeq \frac{H}{k_{\pm}}\to 0\,,\ \ \ k_{+}-k_{-}\to \sigma \,,\
\ \ H|\tau |\to 1\,,\ \ \ \xi _{-}-\xi _{h}\to \xi _{-}r_{h}\,.
\label{Eq/Pg20/1D:dyn_radion_adS4_2_branes}
\end{equation}
In the limit (\ref{Eq/Pg20/1D:dyn_radion_adS4_2_branes}) the relations
between the brane tensions and their positions, eqs.~(\ref{Eq/Pg3/2:dr})
and (\ref{Eq/Pg10/1:dyn_radion_aDs4}), (\ref{Eq/Pg10/2:dyn_radion_aDs4}),
reduce to the well-known fine tuning conditions between the brane tensions
and the bulk cosmological constants,  $\lambda ^{(h)}=4dM^{d}k_{-}$,
$\lambda ^{(v)}=2dM^{d}(k_+-k_{-})$.

\section{Perturbations and gauge}
\label{Section/Pg4/1:dyn_radion/Perturbations}

Let us consider small perturbations of the metric
(\ref{Eq/Pg1/1:dyn_radion})
\[
g_{AB}\to \left\{
\begin{array}{l}
g_{\mu \nu }+\displaystyle\frac{\hat{h}_{\mu \nu
}}{k_{\pm}^{2}\sg^{2}_{\pm}\tau ^{2}}\\
\\
g_{\xi A}+\hat{h}_{\xi A} \,,
\end{array}
\right.
\]
and begin with the coordinate frame $(\hat{x}^{\mu },\hat{\xi })$ in
which the visible brane is placed at $\hat{\xi} =b(\hat{x})$ while the
hidden brane is still at $\hat{\xi} =-\xi _{h}$. We do not assume that
$\hat{h}_{AB}$ is continuous across the brane but the induced metric
\[
\hat{\gamma} _{\mu \nu }=\frac{1}{k_{\pm}^{2}\hat{\tau}
^{2}\mathrm{s}^{2}_{\pm}}\left (\eta_{\mu \nu
}\left(1-2\frac{\mathrm{c}_{\pm}}{\mathrm{s}_{\pm}}b \right)+ \hat{h}_{\mu
\nu } \right)\,,
\]
should be continuous (the first Israel junction condition). Note that
due to the condition (\ref{Eq/Pg2/1:dyn_radion_aDs4}), the coordinates
$(\hat{x}^{\mu },\hat{\xi} )$ continuously cover the whole space.

As an intermediate step, let us demonstrate that there is a gauge in which
the new coordinates $(\tilde{x}^{\mu },\tilde{\xi })$ cover the whole
space, the visible brane is straight and placed at $\tilde{\xi} =0$, the
hidden brane is at $\tilde{\xi} =-\xi _{h}$, and $\tilde{h}_{AB}$ is
continuous. The linear gauge transformation $\delta
h_{AB}(\tilde{X})=\tilde{h}_{AB}(\tilde{X})-\hat{h}_{AB}(\tilde{X})$ of
the metric perturbation under coordinate transformation
$\tilde{X}^{A}=\hat{X}^{A}+\zeta ^{A}$ is
\begin{subequations}
\label{Sub/Pg12/1:dr_arxiv}
\begin{eqnarray}
\delta h_{\xi \xi }&=&-\frac{2}{\sg}(\sg\zeta _{\xi })'\,,
\label{Eqn/Pg3/1:dyn_radion_aDs4}\\
\delta h_{\xi \mu }&=&-\partial _{\mu }\zeta _{\xi
}-\frac{1}{\sg^{2}k^{2}\tau ^{2}}\zeta _{\mu }'\,,
\label{Eqn/Pg3/2:dyn_radion_aDs4}\\
\delta h_{\mu \nu }&=&-\partial _{(\mu }\zeta _{\nu )}+\frac{2}{\tau
}\zeta _{0}\eta _{\mu \nu }-2k^{2}\cg\sg\zeta _{\xi }\eta _{\mu \nu }\,.
\nonumber
\end{eqnarray}
\end{subequations}
Hereafter $(d+1)$-dimensional indices are lowered and raised by $\eta
_{\mu \nu }$, $\xi $-index is lowered and raised by $g_{AB} $, e.g.,
$\zeta _{\xi }=g_{\xi \xi }\zeta ^{\xi }=-\zeta ^{\xi }/k^{2}\sg^{2}$,
prime denotes the derivative with respect to $\xi $, and $a_{(\mu }b_{\nu
)}=a_{\mu }b_{\nu }+a_{\nu }b_{\mu }$.

Let us make the following continuous coordinate transformations:
\begin{eqnarray}
\tilde{\xi} &=&\hat{\xi }-b \cdot \chi (\tilde{\xi })\,, \nonumber \\
\tilde{x}^{\mu }&=&\hat{x}^{\mu }+\zeta ^{\mu }\,, \nonumber
\end{eqnarray}
where $\chi (\tilde{\xi })$ is yet an arbitrary continuous function
satisfying the conditions
\[
\chi (-\xi _{h})=0\,,\ \ \ \ \chi (0)=1.
\]
Then the visible brane is placed at
\[
\tilde{\xi }=0\,,
\]
while the coordinate of the hidden brane is left intact $\tilde{\xi
}=-\xi _{h}$.

In this coordinate frame the jumps of $\tilde{h}_{A\xi }$ across the
visible brane are
\begin{eqnarray}
\triangle \tilde{h}_{\xi \xi }(\tilde{x},\tilde{\xi })&=&\triangle
\hat{h}_{\xi \xi }(\tilde{x},\tilde{\xi
})-2b\triangle\frac{1}{k^{2}\sg}\left(\frac{\chi }{\sg}   \right)'\bigg
|_{\tilde{\xi }=0}\,, \nonumber \\
\triangle \tilde{h}_{\mu \xi }(\tilde{x},\tilde{\xi })&=&\triangle
\hat{h}_{\mu \xi }(\tilde{x},\tilde{\xi
})-\frac{1}{k^{2}\mathrm{s}^{2}\tau ^{2}}\triangle\zeta _{\mu }'
\bigg|_{\tilde{\xi }=0}\,. \nonumber
\end{eqnarray}
We see that the zero jump equations $\triangle \tilde{h}_{A\xi }=0$ can be
satisfied by an appropriate choice of derivatives $\zeta _{\mu }'$ and
$\chi' $ on the brane. Then one has $\triangle \tilde{h}_{\mu \nu }=0$
automatically, due to the first junction condition.

\paragraph{$h_{A\xi }=0$ gauge.} As a final step, we make the second
continuous gauge transformation which gets rid of $h_{A\xi }$ in the whole
space. We write
\begin{equation}
\xi =\tilde{\xi }+\zeta  ^{\xi }\,,\ \ \ \ x^{\mu }=\tilde{x}^{\mu }+\zeta
^{\mu }\,,
\label{Eq/Pg6/1:dyn_radion}
\end{equation}
and require $h_{A\xi }=0$. Then, by making use of eq.
(\ref{Eqn/Pg3/1:dyn_radion_aDs4}), one finds from the condition $h_{\xi
\xi }=0$ that
\[
\zeta  _{\xi }=\frac{1}{2\sg}\int \limits_{0}^{\xi } \sg\tilde{h}_{\xi
\xi }d\xi +\frac{\varepsilon _{\xi }(x)}{\sg}\equiv \zeta _{\xi
}^{(I)}+\frac{\varepsilon _{\xi }(x)}{\sg}\,,
\]
where $\varepsilon_{\xi } (x)$ is in general different in the different
regions. In the ``$-$'' region, $\varepsilon _{\xi }^{-}$ cannot vanish
and is determined by the requirement that the hidden brane is left intact:
$\zeta ^{\xi }(-\xi _{h})=0$, that is
\begin{equation}
\varepsilon ^{-}_{\xi }(x)= \frac{1}{2}\int \limits_{-\xi _{h}}^{0}\sg
\tilde{ h}_{\xi \xi }d\xi .
\label{Eq/Pg6/3:dyn_radion}
\end{equation}
The function $\varepsilon_{\xi } ^{+}(x)$ is determined by the continuity
of $\zeta ^{\xi }$ across the visible brane,
\begin{equation}
k_{-}\varepsilon _{\xi }^{-}=k_{+}\varepsilon _{\xi }^{+}\,.
\label{Eq/Pg7/1:dyn_radion}
\end{equation}
The condition $h_{\mu \xi }=0$ and eq. (\ref{Eqn/Pg3/2:dyn_radion_aDs4})
give
\[
\zeta  _{\mu } =k^{2}\tau ^{2}\int \limits_{0}^{\xi }(\tilde{h}_{\mu
\xi }-\partial _{\mu }\zeta ^{(I)}_{\xi })\sg^{2} d\xi +k^{2}\tau
^{2}\partial _{\mu }\varepsilon _{\xi }\cdot(\mathrm{c}-\cg)+ \varepsilon
_{\mu }(x) \,.
\]
The continuity of $\zeta ^{\mu }$ requires that $\varepsilon _{\mu
}(x)$ is continuous.

Two remarks are in order. First, we note that $\varepsilon _{\mu }(x)$ can
be regarded as a residual gauge transformation,
\begin{equation}
h_{\mu \nu }\to h_{\mu \nu }-\partial _{(\mu }\varepsilon _{\nu
)}+\frac{2}{\tau }\varepsilon _{0}\eta _{\mu \nu }\,,
\label{Eq/Pg7/6:dyn_radion}
\end{equation}
which does not touch the branes and is consistent with the gauge $h_{A\xi
}=0$.

Second, arbitrary functions $\varepsilon _{\xi }^{\pm}(x)$ (which do not
necessarily satisfy the conditions (\ref{Eq/Pg6/3:dyn_radion}),
(\ref{Eq/Pg7/1:dyn_radion})) can be considered as a gauge transformation,
\begin{equation}
h_{\mu \nu }\to h_{\mu \nu }-(\mathrm{c}-\cg)\cdot(2k^{2}\tau ^{2}\partial
_{\mu }\partial _{\nu }\varepsilon _{\xi } +2k^{2}\tau \delta _{(\mu
0}\partial _{\nu )}\varepsilon _{\xi }-2k^{2}\tau \eta _{\mu \nu }\partial
_{\tau }\varepsilon _{\xi })-2k^{2}\cg\eta _{\mu \nu }\varepsilon _{\xi
}\,,
\label{Eq/Pg5/1:dyn_radion_adS4_2_branes}
\end{equation}
which is consistent with the gauge $h_{A\xi }=0$. So, the Einstein
equations in the bulk, being written in the gauge $h_{A\xi }=0$, are
invariant under this transformation. However, these gauge transformations
in general shift the branes. In particular, with the transformations
(\ref{Eq/Pg6/3:dyn_radion}), (\ref{Eq/Pg7/1:dyn_radion}) the hidden brane
is left intact while the new position of the visible brane is determined
at $ \xi^{(v)}=\zeta ^{\xi }(x,0)$ or
\begin{equation}
\xi^{(v)} =f(x)\,,
\label{Eq/Pg5/1:dyn_radion_aDs4}
\end{equation}
where
\[
f(x)=-k^{2}\mathrm{s}\varepsilon _{\xi }
\]
is nothing but the radion. It follows from its definition that the
radion should be continuous across the brane (and it is indeed continuous
due to (\ref{Eq/Pg2/1:dyn_radion_aDs4}) and (\ref{Eq/Pg7/1:dyn_radion})).

In what follows we use continuous coordinates $(x^{\mu },\xi )$ (see
(\ref{Eq/Pg6/1:dyn_radion})), work in the gauge $h_{A\xi }=0$, and place
the hidden brane  at $\xi =-\xi _{h}$, while the position of the visible
brane is given by (\ref{Eq/Pg5/1:dyn_radion_aDs4}). Our purpose is to
derive equation of motion for the radion and study its properties. To this
end we need to find solutions to the perturbed bulk Einstein equations and
Israel junction conditions.

\section{Einstein equations}
\label{Section/Pg8/1:dyn_radion/Einstein equations}
Taking into account
(\ref{Eq/Pg1/4:dyn_radion}) we write the bulk Einstein equations
(\ref{Eq/Pg1/3:dyn_radion}) in the form
\begin{equation}
\mathcal{E}_{AB}\equiv  R_{AB}-k^{2}(d+1)g_{AB}=0\,.
\label{Eq/Pg8/3:dyn_radion}
\end{equation}
In what follows we use the standard helicity decomposition of the metric
perturbation,
\begin{eqnarray}
h_{00}&=&2\Phi \,,\nonumber\\
h_{0i}&=&\partial _{i}Z+Z_{i}\,,\nonumber\\
h_{ij}&=&-2\Psi\delta _{ij} +2\partial _{i}\partial _{j}E+\partial
_{(i}W_{j)}+h_{ij}^{TT}\,, \nonumber
\end{eqnarray}
where $Z_{i}$, $W_{i}$ are transverse and $h_{ij}^{TT}$ is transverse and
traceless
\[
\partial _{i}Z_{i}=\partial _{i}W_{i}=\partial
_{i}h_{ij}^{TT}=h_{ii}^{TT}=0\,.
\]
In terms of these functions the linearized Einstain equations
(\ref{Eq/Pg8/3:dyn_radion}) are
\begin{subequations}
\label{Sub/Pg17/1:dr_arxiv}
\begin{eqnarray}
\delta \mathcal{E}_{00}&=&\partial ^{2}\Phi -\frac{d\dot\Phi}{\tau }
+\frac{2d\Phi }{\tau ^{2}} -d\ddot\Psi +\frac{d\dot\Psi}{\tau }+\partial
^{2}\ddot E-\frac{\partial ^{2}\dot E}{\tau }-\partial ^{2}\dot
Z+\frac{\partial ^{2}Z}{\tau }\nonumber\\
&&+\frac{\hat{O}_{\xi }\Phi }{\tau ^{2}} -\frac{\cg}{2\tau
^{2}\sg}h'\,,
\label{Eqn/Pg7/1:dyn_radion_aDs4}
 \\
\delta \mathcal{E}_{0i}&=&\partial _{i}\left(   -\frac{d-1}{\tau
}\Phi -(d-1)\dot\Psi +\frac{1}{2\tau ^{2}}\hat{O}_{\xi
}Z\right)+\frac{1}{2}\partial ^{2}(Z_{i}-\dot W_{i})+\frac{1}{2\tau
^{2}}\hat{O}_{\xi }Z_{i}\,,
\label{Eqn/Pg17/1:dr_arxiv}\\
\delta \mathcal{E}_{ij}&=&\delta _{ij}\left(\Box\Psi
-\frac{(2d-1)}{\tau }\dot\Psi +\frac{1}{\tau ^{2}}\partial ^{2}(\dot
E-Z)+\frac{\dot\Phi }{\tau }-\frac{2d\Phi }{\tau ^{2}}-\frac{\hat{O}_{\xi
}\Psi  }{\tau ^{2}} +\frac{\cg}{2\tau ^{2}\sg}h'\right)\nonumber\\
&&+\partial _{i}\partial _{j}\left(\dot Z-\frac{(d-1)Z}{\tau
}-(d-2)\Psi -\Phi -\ddot E+\frac{(d-1)\dot E}{\tau }+\frac{\hat{O}_{\xi
}E}{\tau ^{2}}\right)
\label{Eqn/Pg17/2:dr_arxiv}\\
&&+\frac{1}{2}\left(\partial _{(i}\dot Z_{j)}-\frac{(d-1)\partial
_{(i}Z_{j)}}{\tau }-\partial _{(i}\ddot W_{j)}+\frac{(d-1)\partial
_{(i}\dot W_{j)}}{\tau }+\frac{\hat{O}_{\xi }\partial _{(i}W_{j)}}{\tau
^{2}} \right)
\label{Eqn/Pg17/3:dr_arxiv}\\
&&-\frac{1}{2}\left(\Box h_{ij}^{TT}-\frac{(d-1)\dot h_{ij}^{TT}}{\tau
}-\frac{\hat{O}_{\xi }h_{ij}^{TT}}{\tau ^{2}} \right)\,,
\label{Eqn/Pg17/4:dr_arxiv}\\
\delta \mathcal{E}_{0\xi }&=&\left(\partial ^{2}\dot E-\frac{\partial
^{2}E}{\tau }-\frac{\partial ^{2}Z}{2}-\frac{d\Phi }{\tau }-d\dot\Psi
+\frac{d\Psi }{\tau } \right)'\,,
\label{Eqn/Pg18/1:dr_arxiv}\\
\delta \mathcal{E}_{i\xi }&=&\partial _{i}\left(\frac{\dot
Z}{2}-\frac{(d+1)Z}{2\tau }-\Phi -(d-1)\Psi \right)'+\frac{1}{2}\left(\dot
Z_{i}-\frac{(d+1)Z_{i}}{\tau }-\partial ^{2}W_{i} \right)'\,,
\label{Eqn/Pg18/2:dr_arxiv}\\
\delta \mathcal{E}_{\xi \xi }&=&-\sg\left(\frac{h'}{2\sg}   \right)'\,
.
\label{Eqn/Pg7/6:dyn_radion_aDs4}
\end{eqnarray}
\end{subequations}
Hereafter we use the following notations,
\[
h=h_{\mu }^{\mu }= 2\Phi +2d\Psi-2\partial ^{2}E\,,\ \ \ \partial
^{2}=\partial _{i}\partial _{i}\,,\ \ \ \Box=\partial ^{\mu }\partial
_{\mu }=\partial _{0}^{2}-\partial ^{2}\,,
\]
\[
\hat{O}_{\xi }=\partial _{\xi }^{2} -d\frac{\cg }{\sg}\partial _{\xi
}=\sg^{d}\partial _{\xi }\frac{1}{\sg^{d}}\partial _{\xi }\,,
\]
and dot denotes derivative with respect to $\tau $.

\paragraph{Scalars.} The scalar part of eqs. (\ref{Sub/Pg17/1:dr_arxiv})
can be significantly simplified by using variables which are invariant
under the residual gauge transformations (\ref{Eq/Pg7/6:dyn_radion}). Let
us set $\varepsilon _{i}=\partial _{i}\varepsilon $, then the scalar
functions transform as follows,
\[
\delta_{\varepsilon } \Phi=-\tau \partial _{\tau
}\left(\frac{\varepsilon _{0}}{\tau } \right)\,, \ \ \ \delta
_{\varepsilon }\Psi =\frac{\varepsilon _{0}}{\tau }\,, \ \ \ \delta
_{\varepsilon }E=-\varepsilon \,, \ \ \ \delta _{\varepsilon
}Z=-\varepsilon _{0}-\dot\varepsilon \,.
\]
There are two independent gauge-invariant variables.  It is convenient
to use the following pair:
\[
A=\frac{Z-\dot E}{\tau }+\Psi \,,\ \ \ \ B=\dot Z-\ddot E+\Psi -\Phi
\,.
\]
Let us introduce  the  combination,
\[
\frac{U}{2\tau ^{2}}=d\hat{O}_{\tau }A+\partial ^{2}B-\frac{d\dot
B}{\tau }+\frac{d(d+1)B}{\tau ^{2}}\,,
\]
where
\[
\hat{O}_{\tau }=\Box -\frac{(d-1)\partial _{\tau }}{\tau
}-\frac{(d+1)}{\tau ^{2}}\,.
\]
In terms of these variables, the linearized Einstein equations in the
scalar sector take the form
\begin{subequations}
\label{Sub/Pg19/1:dr_arxiv}
\begin{eqnarray}
\delta \mathcal{E}_{00}:&\ \ \ \ &\partial ^{2}A-d\ddot A+\frac{2dA}{\tau
^{2}}+\frac{d\dot B}{\tau }-\frac{2dB}{\tau ^{2}}=-\frac{\hat{O}_{\xi
}\Phi }{\tau ^{2}}+\frac{\cg}{2\tau ^{2}\sg}h'\,,
\label{Eqn/Pg8/1A:dyn_radion_aDs4}\\
\delta \mathcal{E}_{0i}:&\ \ \ \ &-\frac{(d-1)}{\tau }(\tau \dot
A+A-B)=-\frac{\hat{O}_{\xi }Z}{2\tau ^{2}}\,,
\label{Eqn/Pg8/2:dyn_radion_aDs4}\\
\delta \mathcal{E}_{ij}(\delta _{ij}):&\ \ \ \ &\Box A-\frac{2(d-1)\dot
A}{\tau }-\frac{2dA}{\tau ^{2}}-\frac{\dot B}{\tau }+\frac{2dB}{\tau
^{2}}=\frac{\hat{O}_{\xi }\Psi }{\tau ^{2}}-\frac{\cg}{2\tau ^{2}\sg}h'\,,
\label{Eqn/Pg19/1:dr_arxiv}\\
\delta \mathcal{E}_{ij}(\partial _{i}\partial _{j}):&\ \ \ \
&-(d-1)A+B=-\frac{\hat{O}_{\xi }E}{\tau ^{2}}\,,
\label{Eqn/Pg8/1B:dyn_radion_aDs4}\\
\tau ^{d-1}\partial _{\mu }\frac{\delta \mathcal{E}_{\xi }^{\mu }}{\tau
^{d-1}}:&\ \ \ \ &\frac{1}{2\tau ^{2}}(U+dh)'=0\,,
\label{Eqn/Pg8/1C:dyn_radion_aDs4}\\
\delta \mathcal{E}_{\mu }^{\mu }:&\ \ \ \ & \frac{U}{\tau
^{2}}=\frac{\hat{O}_{\xi }h}{2\tau ^{2}}-\frac{\cg(d+1)}{2\tau
^{2}\sg}h'\,.
\label{Eqn/Pg8/1D:dyn_radion_aDs4}
\end{eqnarray}
\end{subequations}
Eqs. (\ref{Eqn/Pg7/6:dyn_radion_aDs4}),
(\ref{Eqn/Pg8/1C:dyn_radion_aDs4}), (\ref{Eqn/Pg8/1D:dyn_radion_aDs4})
yield
\begin{equation}
U=-d\frac{\cg}{\sg}h'\,.
\label{Eq/Pg9/1:dyn_radion_aDs4}
\end{equation}
Combining  eqs. (\ref{Eqn/Pg8/1A:dyn_radion_aDs4}) --
(\ref{Eqn/Pg8/1B:dyn_radion_aDs4}) and their time derivatives and taking
into account (\ref{Eq/Pg9/1:dyn_radion_aDs4}) we finaly obtain the
following set of equations for $A$ and $B$:
\begin{subequations}
\label{Sub/Pg19/2:dr_arxiv}
\begin{eqnarray}
\hat{O}_{\tau }B+\frac{4\dot B}{\tau }+\frac{4B}{\tau
^{2}}&=&\frac{1}{\tau ^{2}}(\hat{O}_{\xi }+d-1)B\,,
\label{Eqn/Pg9/1:dyn_radion_aDs4}\\
\frac{\dot B}{\tau }-\frac{(d-3)B}{\tau ^{2}}-\frac{\partial ^{2}B}{d}
&=&\frac{1}{\tau ^{2}}(\hat{O}_{\xi }+d-1)A\,.
\label{Eq/Pg9/2:dyn_radion_aDs4}
\end{eqnarray}
\end{subequations}

\section{Linearized Israel junction conditions}
\label{Section/Pg12/1:dyn_radion/Linearized Israel conditions}

\subsection{Boundary conditions at $\xi =-\xi _{h}$}

Due to $\mathbb{Z}_{2}$ symmetry, the continuity of the induced metric at
the hidden brane
\[
\gamma _{\mu \nu }^{(h)}=\frac{1}{k_{-}^{2}\mathrm{s}_{h}^{2}\tau
^{2}}(\eta _{\mu \nu }+h_{\mu \nu }(-\xi _{h}))\,,
\]
is trivially satisfied. The jump of the perturbed extrinsic curvature
is given by
\[
-\triangle \delta [\gamma ^{\nu \rho }_{(h)}K_{\rho \mu
}^{(h)}]=-k_{-}\mathrm{s}_{h}h^{\nu' }_{\mu }=0\,.
\]
Thus, one has
\begin{equation}
h_{\mu \nu }'\big|_{\xi =-\xi _{h}}=0\,.
\label{Eq/Pg10/1:dyn_radion_adS4_2_branes}
\end{equation}
This means, in particular, that $ h'\big|_{\xi =-\xi _{h}}=0 $. Together
with eq. (\ref{Eqn/Pg7/6:dyn_radion_aDs4}) this yields
\begin{equation}
h'=0 \ \ \mbox{at}\ \ -\xi _{h}\leq \xi \leq 0\,.
\label{Eq/Pg10/2:dyn_radion_adS4_2_branes}
\end{equation}

\subsection{Junction equations at the visible brane}
\label{Junction equations on the visible brane}

The Israel junction conditions at the visible brane have the form
\begin{subequations}
\label{Sub/Pg21/1:dr_arxiv}
\begin{eqnarray}
\triangle \gamma _{\mu \nu }^{(v)}&=&0\,,
\label{Eqn/Pg9/1A:dyn_radion_aDs4}\\
\triangle K_{\mu \nu }^{(v)}&=&-\sigma \gamma _{\mu \nu }^{(v)}\,.
\label{Eqn/Pg9/2:dyn_radion_aDs4}
\end{eqnarray}
\end{subequations}
The perturbed induced metric on the brane (at $\xi =f(x)$) is given by
\[
\gamma _{\mu \nu }^{(v),\pm}=\frac{1}{k_{\pm}^{2}\tau
^{2}\mathrm{s_{\pm}}^{2}}\left (\eta_{\mu \nu
}\left(1-2\frac{\mathrm{c}_{\pm}}{\mathrm{s}_{\pm}}f \right)+ h_{\mu \nu
}^{\pm} \right)\,,
\]
and the extrinsic curvature is
\[
\gamma ^{\nu \rho }_{(v)}K_{\rho \mu }^{(v)} =-k\mathrm{c}\delta _{\mu
}^{\nu }+k\mathrm{s}\left(\tau ^{2}\partial _{\mu }\partial ^{\nu }f-\tau
(\delta _{\mu }^{\nu }\dot f-\delta _{(\mu 0}\partial ^{\nu )}f)-\delta
_{\mu }^{\nu }f+\frac{h_{\mu }^{\nu '}}{2}\right)\,.
\]
The junction conditions (\ref{Sub/Pg21/1:dr_arxiv}), are satisfied for
the unperturbed background. Hence, for the linearized part we have
\begin{eqnarray}
\triangle\delta \gamma _{\mu \nu }^{(v)}&=&0\,, \nonumber \\
\triangle\delta[\gamma ^{\nu \rho }_{(v)}K_{\rho \mu }^{(v)}]&=&0\,.
\label{Eqn/Pg14/1A:dyn_radion}
\end{eqnarray}
Calculating the trace $K=\gamma ^{\mu \rho }_{(v)}K_{ \mu \rho }^{(v)}$ we
get
\begin{equation}
K=-k\mathrm{c}(d+1)+k\mathrm{s}\tau ^{2}\left(\hat{O}_{\tau
}f+\frac{h'}{2\tau ^{2}} \right)\,.
\label{Eq/Pg10/3:dyn_radion_aDs4}
\end{equation}
By making use the Gauss-Codazzi relation
\begin{equation}
2G_{AB}n^{A}n^{B}=k_{\pm}^{2}d(d+1)=^{(d+1)}R-(K_{\mu \nu
}^{\pm(v)}K_{\pm(v)}^{\mu \nu }-K_{\pm}^{2})\,,
\label{Eq/Pg15/1B:dyn_radion}
\end{equation}
where $G_{AB}$ is the Einstein tensor, $^{(d+1)}R$ is the curvature scalar
on the brane, and $K^{\mu \nu }_{(v)}\equiv \gamma ^{\mu \rho
}_{(v)}\gamma ^{\nu \lambda }_{(v)}K_{\rho \lambda }^{(v)}$, one finds
\begin{equation}
\triangle\delta (K_{\mu \nu }^{(v)}K^{\mu \nu }_{(v)}-K^{2})=0\,.
\label{Eq/Pg15/1C:dyn_radion}
\end{equation}
Due to the fact that the background extrinsic curvature is proportional to
$\eta _{\mu \nu }$ (cf. (\ref{Eq/Pg3/4:dr})), it is straightforward to
check that eq.~(\ref{Eq/Pg15/1C:dyn_radion}) takes the form
\[
-\triangle\left[2dk\mathrm{c}\cdot\delta K \right]=0\,.
\]
Together with eq. (\ref{Eqn/Pg14/1A:dyn_radion}) this leads to the
equation
\[
\delta K_{\pm} =0\,,
\]
and, therefore,
\begin{equation}
\hat{O}_{\tau }f=0\,,
\label{Eq/Pg10/4:dyn_radion_aDs4}
\end{equation}
where we have used (\ref{Eq/Pg10/2:dyn_radion_adS4_2_branes}). This is the
desired radion equation of motion.

Besides that, the junction conditions yield
\begin{eqnarray}
\triangle h_{\mu \nu }&=&2\triangle\frac{ \mathrm{c}}{\mathrm{s}}\eta
_{\mu \nu }f=2\frac{\sigma }{H}\eta _{\mu \nu }f\,,
\label{Eqn/Pg11/1:dyn_radion_aDs4}\\
\triangle h_{\mu \nu }'&=&0\,. \nonumber
\end{eqnarray}
From the latter equation and eqs. (\ref{Eqn/Pg7/6:dyn_radion_aDs4}),
(\ref{Eqn/Pg8/1D:dyn_radion_aDs4}),
(\ref{Eq/Pg10/2:dyn_radion_adS4_2_branes}) we find
\begin{equation}
h'=0\,,\ \ \ \ U=0
\label{Eq/Pg12/1:dyn_radion_adS4_2_branes}
\end{equation}
in the whole space.

The condition (\ref{Eqn/Pg11/1:dyn_radion_aDs4}) translates into
\begin{equation}
\triangle \Phi =\triangle \Psi =\triangle A= \frac{\sigma }{H}f\,,
\label{Eq/Pg11/1:dyn_radion_aDs4}
\end{equation}
while other functions characterizing the metric perturbations, as well as
all first derivatives of $h_{\mu \nu }$ with respect to $\xi $ are
continuous across the brane.

\section{Solutions}
\label{Section/Pg11/1:dyn_radion_aDs4/The Einstein equations solution}
\subsection{Scalar sector}
\label{Subsec/Pg11/1:dyn_radion_aDs4/Scalar sector}
Now we are ready to solve the linearized Einstein equations. We begin with
eq.~(\ref{Eqn/Pg9/1:dyn_radion_aDs4}). The variables separate, so the
modes have the form
\[
B_{\kappa  }(x,\xi )=b_{\kappa  }(x)\beta _{\kappa }(\xi )\,,
\]
where $\beta _{\kappa  }$ are normalizable (since $B$ is gauge
invariant),
\[
\int \limits_{-\xi _{h}}^{\infty }\frac{d\xi }{(k\sg)^{d}}|\beta
_{\kappa }|^{2}<\infty\,,
\]
and continuous together with their derivatives across the brane (see
eq.~(\ref{Eq/Pg11/1:dyn_radion_aDs4})):
\[
\triangle \beta_{\kappa  }(0) =\triangle\beta _{\kappa  }'(0)=0\,.
\]
They are solutions to the eigenvalue equation
\begin{equation}
(\hat{O}_{\xi }+d-1)\beta _{\kappa }=-\nu \beta _{\kappa }\, .
\label{Eq/Pg12/2:dyn_radion_aDs4}
\end{equation}
Explicitly,
\begin{equation}
\sg^{d}\frac{\partial }{\partial \xi }\frac{1}{\sg^{d}}\frac{\partial
}{\partial \xi }\beta _{\kappa } +\left(\frac{d^{2}}{4}-\kappa
^{2}\right)\beta _{\kappa }=0
\label{Eq/Pg16/3:dyn_radion_adS4}
\end{equation}
with
\[
\kappa =\frac{\sqrt{(d-2)^{2}-4\nu }}{2}\;.
\]
We show in Appendix \ref{Appendix Spectrum} that there is one constant
discrete mode in the spectrum with $\kappa =d/2$ ($\nu =1-d$),
\begin{equation}
\beta _{\frac{d}{2}}(\xi )=\mathrm{const}\,.
\label{Eq/Pg13/1:dyn_radion_adS4_2_branes}
\end{equation}
For $k_{-}>0$ it is localized near the hidden brane. However, as we
discuss later on, this mode does not generate a solution to the complete
set of the Einstein equations~(\ref{Sub/Pg19/1:dr_arxiv}), so the
corresponding metric perturbations are, in fact, absent. The rest of the
spectrum is continuous and starts from zero: $\kappa ^{2}\leq 0$ ($\nu
\geq (d-2)^{2}/4$). The $x^{\mu }$-dependent parts
\[
b_{\kappa  }(x) =b_{\kappa  }(\tau
,\mathbf{p})\mathrm{e}^{i\mathbf{px}}\,,
\]
satisfy  the following equation,
\begin{equation}
\left(\partial _{\tau }^{2}+p^{2}-\frac{(d-5)\partial _{\tau }}{\tau
}-\frac{(d-3-\nu )}{\tau ^{2}}\right)b_{\kappa  }(\tau ,\mathbf{p})=0\,.
\label{Eq/Pg13/4:dyn_radion_aDs4}
\end{equation}

Let us now consider eq.~(\ref{Eq/Pg9/2:dyn_radion_aDs4}). For
non-vanishing left hand side this equation immediately yields
\begin{equation}
A_{\kappa  }(\tau ,\mathbf{p},\xi )=a_{\kappa  }(\tau ,\mathbf{p})\beta
_{\kappa  }(\xi )\,,
\label{Eq/Pg14/2:dyn_radion_aDs4}
\end{equation}
with
\begin{equation}
a_{\kappa  }(\tau ,\mathbf{p})=\frac{1}{d\nu }\left(d(d-3)-d\tau \partial
_{\tau }-p^{2} \right)b_{\kappa  }(\tau ,\mathbf{p})\,.
\label{Eq/Pg14/3:dyn_radion_aDs4}
\end{equation}
There is an imporant subtlety here. The modes
(\ref{Eq/Pg14/2:dyn_radion_aDs4}) are continuous across the visible brane
and hence contribute to the continuous part of the function $A$ only. This
continuous part of $A$ satisfies eq.~(\ref{Eq/Pg11/1:dyn_radion_aDs4})
with vanishing right hand side. To satisfy
eq.~(\ref{Eq/Pg11/1:dyn_radion_aDs4}) with non vanishing right hand side,
we note that the operator $\hat{O}_{\xi } +d-1$ has yet another zero mode
(in addition to (\ref{Eq/Pg13/1:dyn_radion_adS4_2_branes})) when it acts
in the space of discontinuous functions. In that case both sides of
eq.~(\ref{Eq/Pg9/2:dyn_radion_aDs4}) are equal to zero, and hence the
relations (\ref{Eq/Pg14/2:dyn_radion_aDs4}),
(\ref{Eq/Pg14/3:dyn_radion_aDs4}) are no longer valid.

Thus, we search for the solution of the form
\begin{equation}
A_{\frac{d-2}{2}}(x,\xi )=f(x)\beta _{\frac{d-2}{2}}(\xi )\,, \ \  \ \
B_{\frac{d-2}{2}}(x,\xi )=0\,,
\label{Eq/Pg15/1:dyn_radion_aDs4}
\end{equation}
where the second equality follows from the fact that
eqs. (\ref{Sub/Pg19/2:dr_arxiv}) do not admit non-trivial solution for $B$
in the case of vanishing right hand side of
eq.~(\ref{Eq/Pg9/2:dyn_radion_aDs4}). The function $\beta
_{\frac{d-2}{2}}(\xi )$ must obey eq.~(\ref{Eq/Pg12/2:dyn_radion_aDs4}) in
both ``+'' and ``-'' regions and has the jump at the visible brane
\begin{equation}
\triangle \beta _{\frac{d-2}{2}} = \frac{\sigma}{H} \; .
\label{jul8-16-1}
\end{equation}
The boundary condition at the hidden brane follows from
(\ref{Eq/Pg10/1:dyn_radion_adS4_2_branes}):
\begin{equation}
\beta^\prime_{\frac{d-2}{2}} (-\xi_h)= 0 \; .
\label{jul8-16-2}
\end{equation}
To construct the new zero mode we note that  two linear independent
solutions to eq.~(\ref{Eq/Pg16/3:dyn_radion_adS4}) with $\kappa =(d-2)/2$
($\nu=0$) are
\begin{subequations}
\label{Eq/Pg25/1:dr_arxiv}
\begin{eqnarray}
\beta _{\frac{d-2}{2}}^{(1)}&=&\cg^{d-1} \left(\frac{\sg}{\cg}
\right)^{d+1}F\left(1,\frac{3}{2}; \frac{d+3}{2}; \frac{\sg^2}{\cg^2}
\right)\,,
\label{Eqn/Pg12/1:dyn_radion_aDs4}\\
\beta _{\frac{d-2}{2}}^{>}&=&\cg \;,
\label{Eqn/Pg15/1:dyn_radion_adS4_2_branes}
\end{eqnarray}
\end{subequations}
where $F$ is  the hypergeometric function. At large $\xi $, $\beta
_{\frac{d-2}{2}}^{(1)}$ grows as $\mathrm{e}^{(d-1)\xi }$ and hence it
cannot be used in the ``+'' region. In contrast, the second solution
$\beta _{\frac{d-2}{2}}^{>}$ is sutable at large $\xi $. In the ``$-$''
region the following linear combination of (\ref{Eq/Pg25/1:dr_arxiv})
satisfies (\ref{jul8-16-2}):
\begin{equation}
\beta _{\frac{d-2}{2}}^{<} (\xi)=\beta _{\frac{d-2}{2}}^{(1)}(\xi
)-\cg_{-} \frac{\beta _{\frac{d-2}{2}}^{(1)'}(-\xi _{h})}{\mathrm{s}_{h}}
\label{Eq/Pg15/1:dyn_radion_adS4_2_branes}
\; .
\end{equation}
By making use of the boundary condition (\ref{jul8-16-1}) at $\xi =0$ we
finaly obtain
\begin{equation}
\beta _{\frac{d-2}{2}}(\xi )=\left\{
\begin{array}{l}
\displaystyle{\frac{\sigma
}{H}\left(\mathrm{s}_{+}\left[\frac{\mathrm{c_{+}}}{\mathrm{s_{+}}}-
\frac{\beta _{\frac{d-2}{2}}^{<}(0)}{\beta _{\frac{d-2}{2}}^{<'}(0)}
\right] \right)^{-1}\cg_{+}}\ \ ~~\mbox{at}\ \ \xi >0\,,\\
\\
\displaystyle{\frac{\sigma }{H}\left(\beta
_{\frac{d-2}{2}}^{<'}(0)\left[\frac{\mathrm{c_{+}}}{\mathrm{s_{+}}}-
\frac{\beta _{\frac{d-2}{2}}^{<}(0)}{\beta _{\frac{d-2}{2}}^{<'}(0)}
\right] \right)^{-1}\beta _{\frac{d-2}{2}}^{<}(\xi )}\ \ ~~\mbox{at}\ \
\xi <0\,,
\end{array}
\right.
\label{Eq/Pg14/5:dyn_radion_aDs4}
\end{equation}
In both of these formulas, $\beta _{\frac{d-2}{2}}^{<}(0)$ and $\beta
_{\frac{d-2}{2}}^{<'}(0)$ are the limiting values in the ``-'' region. To
end up with the analysis of the zero mode, we note that the Wronskian
$\mathcal{ W}(f_{1},f_{2})=f_{1}f_{2}'-f_{1}'f_{2}$ of the functions
(\ref{Eq/Pg25/1:dr_arxiv}) is
\begin{equation}
\mathcal{W}(\cg, \beta _{\frac{d-2}{2}}^{(1)})=\mathcal{ W}(\cg,\beta
_{\frac{d-2}{2}}^{<})=(d+1)\sg^{d}\,.
\label{Eq/Pg16/1A:dyn_radion_adS4_2_branes}
\end{equation}
Note also that the terms  proportional to $\cg_{\pm}$ in
(\ref{Eq/Pg14/5:dyn_radion_aDs4}) correspond to the gauge transformation
that  preserves $h_{\xi \xi }=0$ (cf. eq.
(\ref{Eq/Pg5/1:dyn_radion_adS4_2_branes})). In particular, the radion is
pure gauge in the ``+'' region outside the visible brane, i.e., the
non-trivial part of its wave function is concentrated on and between the
branes.

Let us now come back to the constant mode
(\ref{Eq/Pg13/1:dyn_radion_adS4_2_branes}) and consider eqs.
(\ref{Eqn/Pg8/2:dyn_radion_aDs4}), (\ref{Eqn/Pg8/1B:dyn_radion_aDs4}).
These equations can be viewed as inhomogeneous equations for $Z$ and $E$,
respectively. Recall that the operator $\hat{O}_{\xi }$ has exactly one
zero mode $\beta _{d/2}$. The necessary condition for the existence of
solutions to eqs. (\ref{Eqn/Pg8/2:dyn_radion_aDs4}),
(\ref{Eqn/Pg8/1B:dyn_radion_aDs4}) is the orthogonality of the
inhomogeneity to this mode, and it cannot be satisfied if $A$ and/or $B$
contain  contributions proportional to $\beta _{d/2}$. Thus, we are forced
to conclude that $B$ contains  the continuous part of the spectrum of
(\ref{Eq/Pg12/2:dyn_radion_aDs4}) only. On the contrary, the
(discontinuous) zero mode $\beta _{\frac{d-2}{2}}$, contributing to $A$,
is orthogonal to $\beta _{d/2}$. Indeed, by making use of
(\ref{Eq/Pg12/2:dyn_radion_aDs4}),
(\ref{Eq/Pg13/1:dyn_radion_adS4_2_branes}), integrating by parts and
taking into account the boundary conditions at $-\xi _{h}$ and at
infinity, we write
\[
\langle \beta _{\frac{d}{2}}|\beta _{\frac{d-2}{2}}\rangle =\beta
_{\frac{d}{2}}\cdot\int \limits_{-\xi _{h}}^{\infty }\frac{d\xi
}{(k\sg)^{d}}\frac{\hat{O}_{\xi }\beta _{\frac{d-2}{2}}}{1-d} =\beta
_{\frac{d}{2}}\cdot \frac{\triangle \beta_{\frac{d-2}{2}}
'(0)}{k\mathrm{s}(d-1)}=0\,.
\]
The last point to check is that $U$ and $h'$  vanish, eq.
(\ref{Eq/Pg12/1:dyn_radion_adS4_2_branes}). Using eqs.
(\ref{Eq/Pg13/4:dyn_radion_aDs4}), (\ref{Eq/Pg14/3:dyn_radion_aDs4}) one
directly finds that eq. (\ref{Eq/Pg12/1:dyn_radion_adS4_2_branes}) indeed
holds for the modes with $\nu \neq 0$. For the zero mode
(\ref{Eq/Pg15/1:dyn_radion_aDs4}),
eq.~(\ref{Eq/Pg12/1:dyn_radion_adS4_2_branes}) is satisfied due to the
radion equation of motion (\ref{Eq/Pg10/4:dyn_radion_aDs4}).

Explicit expressions for the metric components induced by the radion can
be found by making use eqs. (\ref{Eqn/Pg8/1A:dyn_radion_aDs4}) --
(\ref{Eqn/Pg8/1B:dyn_radion_aDs4}). Let $Q(\xi )$ be a continuous solution
to the equation
\[
\hat{O}_{\xi }Q=\beta _{\frac{d-2}{2}}(\xi )
\]
with boundary conditions
\[
Q'(-\xi _{h})=0\,, \ \ \ \frac{Q(\xi)}{\sg_{+}^{d/2}} \bigg |_{\xi \to
\infty }\to 0\,.
\]
Explicitly,
\[
Q(\xi )=\frac{\beta _{\frac{d-2}{2}}(\xi)}{1-d} +\Theta (-\xi )
\frac{\sigma }{H(1-d)} +\mbox{const}\,,
\]
where $\Theta $ is the step function; the last constant term cannot be
fixed and corresponds to the residual gauge transformation
(\ref{Eq/Pg7/6:dyn_radion}). Then
\begin{eqnarray}
\Phi &=& f\beta _{\frac{d-2}{2}}+(d-1)(\tau ^{2}\ddot f +\tau \dot f)\cdot
Q\,, \nonumber \\
\Psi  &=&f\beta _{\frac{d-2}{2}}-(d-1)\tau \dot f\cdot Q\,, \nonumber \\
E&=&(d-1)\tau ^{2}f\cdot Q\,, \nonumber \\
Z&=&2(d-1)(\tau ^{2}\dot f+\tau f)\cdot Q\,. \nonumber
\end{eqnarray}

\subsection{Vector sector}
\label{Subsec/Pg16/1:dyn_radion_aDs4/Vsector}

Let us introduce the following gauge invariant variable:
\[
V_{i}=Z_{i}-\dot W_{i} \,.
\]
Then the Einstein equations in the vector sector are
\begin{subequations}
\label{Sub/Pg29/1:dr_arxiv}
\begin{eqnarray}
&&\Box V_{i}-\frac{(d-3)\dot V_{i}}{\tau }=\frac{1}{\tau
^{2}}(\hat{O}_{\xi }+d-1)V_{i}\,,
\label{Eqn/Pg16/3:dyn_radion_aDs4}\\
&&\partial ^{2}V_{i}=-\frac{\hat{O}_{\xi }Z_{i}}{\tau
^{2}}\,,
\label{Eqn/Pg17/1:dyn_radion_adS4_2_branes}\\
&&\left(\dot Z_{i}-\frac{(d+1)Z_{i}}{\tau }-\partial ^{2}W_{i}
\right)'=0
\label{Eqn/Pg17/2:dyn_radion_adS4_2_branes}
\,.
\end{eqnarray}
\end{subequations}
The situation is reminiscent of that in the scalar sector. Any solution to
eq. (\ref{Eqn/Pg16/3:dyn_radion_aDs4}) can be decomposed in eigenfunctions
$\beta _{\kappa }(\xi )$. However, the contribution from the localized
mode $\beta _{d/2}$ vanishes due to the second equation
(\ref{Eqn/Pg17/1:dyn_radion_adS4_2_branes}): $V_{i}$ should be orthogonal
to $\beta _{d/2}$. Then the validity of the third equation
(\ref{Eqn/Pg17/2:dyn_radion_adS4_2_branes}) can be directly verified.
Thus, all vector modes belong to the continuous part of the spectrum of
the operator (\ref{Eq/Pg12/2:dyn_radion_aDs4}), and hence they are
delocalized.

\subsection{Tensor sector}
\label{Subsec/Pg18/1:dyn_radion_adS4_2_branes/Tensor sector}

The only equation in the tensor sector is
\[
\Box h_{ij}^{TT}-\frac{(d-1)\dot h_{ij}^{TT}}{\tau
}+\frac{(d-1)h_{ij}^{TT}}{\tau ^{2}}= \frac{1}{\tau ^{2}}(\hat{O}_{\xi
}+d-1)h_{ij}^{TT}\,.
\]
Therefore, there are no conditions eliminating the discrete mode $\beta
_{d/2}$ which in the case $k_{-}>0$ is localized near the hidden brane. By
writing
\[
h_{(\kappa )ij}^{TT}(X)=\beta _{\kappa }(\xi )\cdot
e_{ij}\cdot\mathcal{ H}_{\kappa }(\tau
,\mathbf{p})\mathrm{e}^{i\mathbf{px}}\,,
\]
where $e_{ij}$ is constant transverse-traceless polarization tensor,
one finds the equation for $\mathcal{ H}_{d/2}$:
\[
\left(  \partial ^{2}_{\tau }+p^{2}-\frac{(d-1)\partial _{\tau }}{\tau
} \right) \mathcal{ H}_{\frac{d}{2}}=0\,,
\]
which is precisely the equation for the graviton perturbations in the
de Sitter $(d+1)$-dimensional Universe. The negative frequency solution to
this equation at $\tau <0$ is
\[
\mathcal{ H}_{\frac{d}{2}}=(-p\tau )^{d/2}H_{\frac{d}{2}}^{(1)}(-p\tau
)\, ,
\]
where $H_{\kappa}^{(1)}$ is the Hankel function. This solution leads to
the flat power spectrum for the tensor modes.

\section{Effective action for the light modes}
\label{Section/Pg19/1:dyn_radion_adS4_2_branes/Light modes effective
action}

\subsection{Effective action for the radion}
\label{Subsec/Effective action for the radion}

In this section we calculate the quadratic effective action for the radion
and the graviton zero mode.   To this end, we make use of the first
variation of the action (\ref{Eq/Pg1/2:dyn_radion}). We begin with the
radion. A subtlety is that in our gauge the action depends on the radion
not only through the metric components but also through the visible brane
position $\xi =f(x)$. To get around this difficulty we perform a gauge
transformation that puts the visible brane at the origin, straightens it,
but does not touch the hidden brane:
\[
\xi \to \xi -f(x)\chi (\xi )\,,\ \ \ \ \chi (0)=1\,,\ \ \ \chi (-\xi
_{h})=0\,,
\]
where $\chi $ is continuous together with its first derivative at $\xi
=0$. This gauge transformation leads to  nonvanishing components
$\tilde{h}_{\xi A}$, in particular,
\begin{equation}
\tilde{h}_{\xi \xi }=-\frac{2f}{k^{2}\sg}\left(\frac{\chi }{\sg}
\right)'\,.
\label{Eq/Pg20/1A:dyn_radion_adS4_2_branes}
\end{equation}
On the other hand, one can  keep the conditions $\tilde{h}_{\xi \mu }=0$
by making another gauge transformation $x^{\mu }\to x^{\mu }+\zeta ^{\mu
}$ with
\[
\zeta _{\mu }=-\tau ^{2}\partial _{\mu }f\int \limits_{}^{\xi }\chi
d\xi \,.
\]
Then the $(\mu \nu )$-components of the metric perturbations become
\[
\tilde{h}_{\mu \nu }=h_{\mu \nu }+2(\tau ^{2}\partial _{\mu }\partial
_{\nu }f+\tau \delta _{(\mu 0}\partial _{\nu )}f-2\tau \dot f \eta _{\mu
\nu })\int \limits_{}^{\xi }\chi d\xi -2\frac{\cg}{\sg}f\chi \eta _{\mu
\nu }\,.
\]
It worth noting that $\tilde{h}_{\mu \nu }$ is continuous across the
visible brane while the jump of its derivative is
\[
\triangle \tilde{h}_{\mu \nu }'=2\frac{ k^{2}_{+}-k^{2}_{-}}{H^{2}}f\,.
\]

Now, the quadratic action for the radion is
\begin{equation}
S_{f}=-\frac{1}{2}\int
\limits_{}^{}dX^{d+2}\sqrt{g}\tilde{h}_{(f)}^{AB}\delta
E_{AB}[\tilde{h}_{(f)}]\,,
\label{Eq/Pg20/5:dyn_radion_adS4_2_branes}
\end{equation}
where the subscript $(f)$ means that we take into account only the part of
perturbations depending on the (off-shell) radion, and the tensor $\delta
E_{AB}$ is the linear part of the variation of the action
(\ref{Eq/Pg1/2:dyn_radion}),
\[
\delta E_{AB} = M^{d}\delta \left(R_{AB} -\frac{1}{2} g_{AB} R
+k^{2}d(d+1)g_{AB} \right) \; .
\]
As in the static case~\cite{PRZ, Dubovsky:2003pn}, the only
non-vanishing component is
\begin{equation}
\delta E_{\xi \xi }=M^{d}\frac{d}{d-1}\sg\left(\frac{\beta
_{\frac{d-2}{2}}'}{\sg} \right)'\tau ^{2}\hat{O}_{\tau }f
=M^{d}\frac{d}{\sg}\mathcal{ W}(\cg,\beta _{\frac{d-2}{2}})\tau
^{2}\hat{O}_{\tau }f\,,
\label{Eq/Pg20/6:dyn_radion_adS4_2_branes}
\end{equation}
where  we have used eq. (\ref{Eq/Pg12/2:dyn_radion_aDs4}) to obtain the
last equality.

\begin{figure}[t]
\begin{center}
\includegraphics[
height=100mm,
]{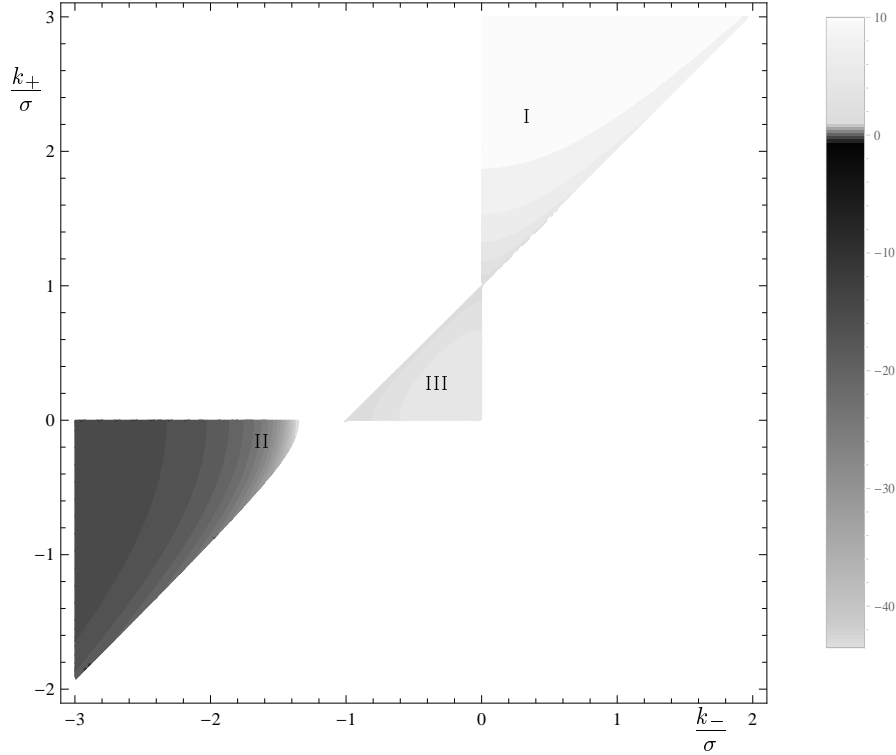}
\end{center}
\caption{Contour plot of $\mathcal{ P}H^{2}/(|\sigma |M^{d})$ as a
function of $k_{-}/\sigma$, $k_{+}/\sigma$ at $\xi _{h}=0.3$. The regions
I--III correspond to the allowed regions in the parameter space discussed
in Sec.~\ref{Section/Pg1/1:dyn_radion/The Setup}. In the region II (more
dark), which corresponds to the negative visible brane tension, the radion
is a ghost.
\label{Fig/Pg23/1:dyn_radion_adS4_2_branes}
}
\end{figure}

Due to eq. (\ref{Eq/Pg16/1A:dyn_radion_adS4_2_branes}), upon substituting
eqs. (\ref{Eq/Pg20/1A:dyn_radion_adS4_2_branes}) and
(\ref{Eq/Pg20/6:dyn_radion_adS4_2_branes}) into
(\ref{Eq/Pg20/5:dyn_radion_adS4_2_branes}), we find that the $\xi
$-dependent part of the integrand of
(\ref{Eq/Pg20/5:dyn_radion_adS4_2_branes})  is total derivative:
\[
\frac{1}{|k^{d}\sg^{d-2}|}\left(\frac{\beta _{\frac{d-2}{2}}'}{\sg}
\right)'\left(\frac{\chi }{\sg} \right)' =\Theta (-\xi
)(d-1)\left(\frac{\chi }{\sg_{-}} \frac{\mathcal{ W}(\cg_{-},\beta
_{\frac{d-2}{2}})}{|k_{-}\sg_{-}|^{d}} \right)'\,.
\]
Taking into account that we work on the full $\xi$ axis  with
$\mathbb{Z}_{2}$ identification, and introducing a new field
\begin{equation}
\phi =\frac{\sqrt{|\mathcal{ P}|}f(x)}{|H\tau| ^{\frac{d-1}{2}}}\,,
\label{Eq/Pg21/1B:dyn_radion_adS4_2_branes}
\end{equation}
where
\begin{equation}
\mathcal{ P}= 4d(d+1)\frac{M^{d}}{H} \left((d+1)
\frac{k_{+}\mathrm{c}_{+}H}{k_{-}\mathrm{c}_{-}\sigma } +\frac{\beta
_{\frac{d-2}{2}}^{<}(0)}{\mathrm{c}_{-}\mathrm{s}_{-}^{d-2}}\right)^{-1}
\; ,
\label{Eq/Pg21/1D:dyn_radion_adS4_2_branes}
\end{equation}
we finaly arrive at the radion effective action
 \begin{equation}
S_{f}=\mbox{sign}(\mathcal{ P})\int \limits_{}^{}d^{d+1}x\left(
\frac{1}{2}(\partial _{\mu }\phi )^{2}+\frac{(d+1)(d+3)}{8\tau ^{2}}\phi
^{2}\right)\, .
 \label{Eq/Pg21/1C:dyn_radion_adS4_2_branes}
\end{equation}
The normalization factor (\ref{Eq/Pg21/1D:dyn_radion_adS4_2_branes}) is
obtained by making use of eqs. (\ref{Eq/Pg14/5:dyn_radion_aDs4}),
(\ref{Eq/Pg16/1A:dyn_radion_adS4_2_branes}). It worth  noting that the
radion is a ghost at $\sigma <0$ (see
Fig.~\ref{Fig/Pg23/1:dyn_radion_adS4_2_branes}).

In the static limit (\ref{Eq/Pg20/1D:dyn_radion_adS4_2_branes}) one has
\[
\mathcal{ P} = 4dM^{d}\frac{k_{-}}{H^{2}r_{h}^{d-1}}
\left(\frac{k_{+}}{r_{h}^{d-1}\triangle k} -1 \right)^{-1}\,,
\]
which, modulo notations, coincides with the result of Refs.~\cite{PRZ,
Dubovsky:2003pn}.

Up to the sign of $\mathcal{ P}$, the action
(\ref{Eq/Pg21/1C:dyn_radion_adS4_2_branes}) coincides with the action for
perturbations about a time-dependent background in a $(d+1)$-dimensional
classical conformal theory. The latter theory is described by the action
\begin{eqnarray}
S_{\varphi }&=&\int \limits_{}^{}d^{d+1}x\sqrt{-\gamma
}\left[\frac{1}{2}\gamma ^{\mu \nu }\partial _{\mu }\varphi \partial _{\nu
}\varphi+\frac{^{(d+1)}R(d-1)}{8d}\varphi ^{2}-\frac{(-\lambda
^{2})}{2}\varphi ^{\frac{2(d+1)}{d-1}}\right] \nonumber\\
&=&\int \limits_{}^{}d^{d+1}x\left[\frac{1}{2}(\partial _{\mu
}\tilde\varphi)^{2} -\frac{(-\lambda ^{2})}{2}\tilde\varphi
^{\frac{2(d+1)}{d-1}}\right]\,,
\label{Eq/Pg21/1E:dyn_radion_adS4_2_branes}
\end{eqnarray}
with $\gamma _{\mu \nu }=a^{2}\eta _{\mu \nu }$, $\tilde{\varphi
}=a^{\frac{d-1}{2}}\varphi $, while the background time-dependent solution
is, at $\tau <0$,
 \[
\tilde\varphi _{c}= \left(\frac{d-1}{2\lambda }
\right)^{\frac{d-1}{2}}\frac{1}{(-\tau )^{\frac{d-1}{2}}} \;.
\]
Modulo the replacement of the real field $\varphi $ by complex one,
eq.~(\ref{Eq/Pg21/1E:dyn_radion_adS4_2_branes}) is precisely the action
considered in the context of (pseudo)conformal Universe
model~\cite{Rubakov:2009np, Libanov:2015iwa}.

The equation of motion for the canonicaly normalized radion $\phi $
obtained from the action (\ref{Eq/Pg21/1C:dyn_radion_adS4_2_branes}), is
\[
\Box \phi -\frac{(d+1)(d+3)}{4\tau ^{2}}\phi =0\,.
\]
Its negative frequency solution that tends to properly normalized mode
of free quantum field as $p\tau \to -\infty $ is
\[
\phi =\frac{\sqrt{-\tau }}{2^{\frac{d+2}{2}}\pi^{\frac{d-1}{2}}
}\mathrm{e}^{\frac{i\pi }{4}(d-5)}H^{(1)}_{\frac{d+2}{2}}(-p\tau
)\mathrm{e}^{i\mathbf{px}} \,.
\]
At late times, when $-p\tau \ll 1$, one has
\[
\phi =\mathrm{e}^{i\mathbf{kx}}\cdot\mathrm{e}^{\frac{i\pi
}{4}(d+1)}\Gamma \left(\frac{d+2}{2}   \right)\frac{1}{\pi
^{\frac{d+1}{2}}p^{\frac{d+2}{2}}(-\tau )^{\frac{d+1}{2}}}\,,
\]
what leads to the red power spectrum (\ref{Eq/Pg1/2:dr}).

\subsection{Radion-matter coupling}
\label{Subsec/Radion to matter coupling}
Let $T_{AB}^{\pm}$ and $\mathcal{
T}_{\mu \nu }$ be the energy-momentum tensors of matter residing in the
bulk and on the visible brane, respectively. $T_{AB}^{\pm}$ does not
include contributions from the bulk cosmological constants and can be, in
general, different in the different regions, while $\mathcal{ T}_{\mu \nu
}$ does not include the brane tension. We assume that the matter
energy-momentum tensors are small and treat them as perturbations. For
simplicity we also assume that there is no matter residing on the hidden
brane. To derive the radion equation of motion in the presence of matter
we note that in this case eqs. (\ref{Eq/Pg15/1B:dyn_radion}) and
(\ref{Eqn/Pg9/2:dyn_radion_aDs4}) take the form
\[
^{(d+1)}R-(K_{\mu \nu }^{\pm(v)}K_{\pm(v)}^{\mu \nu
}-K_{\pm}^{2})-k_{\pm}^{2}d(d+1)-\frac{n_{A}n_{B}T^{AB}_{\pm}}{M^{d}}=0\,,
\]
\[
\triangle K_{\mu \nu }^{(v)}=-\sigma \gamma _{\mu \nu }^{(v)}
+\frac{1}{2M^{d}}\left(\mathcal{ T}_{\mu \nu }-\frac{\gamma _{\mu \nu
}\gamma ^{\lambda \rho }\mathcal{ T}_{\lambda \rho }}{d} \right)\,.
\]
To the leading order in perturbations about the source-free background,
one has from these equations
\[
^{(d+1)}R+d(d+1)\mathrm{s}_{\pm}^{2}k^{2}_{\pm}-2d\mathrm{c}_{\pm}k_{\pm}\delta
K_{\pm}-\frac{H^{2}T_{\xi \xi }^{\pm}}{M^{d}}=0\,,
\]
\[
\triangle \delta K=-\frac{H^{2}\tau ^{2}\mathcal{ T}}{2dM^{2}}\,,
\]
where $\mathcal{ T}\equiv \eta ^{\mu \nu }\mathcal{ T}_{\mu \nu }$. We
actually have three equations, which can be used to find the induced
scalar curvature $^{(d+1)}R$ and the values of $\delta K_{\pm}$ on both
sides of the visible brane. The result for $\delta K_{\pm}$ is
\begin{equation}
\delta K_{\pm}=\frac{1}{2dM^{d}\sigma
}\left(\mathrm{c}_{\mp}k_{\mp}H^{2}\tau ^{2}\mathcal{ T}-H^{2}\triangle
T_{\xi \xi } \right)\,.
\label{Eq/Pg24/2A:dyn_radion_adS4_2_branes}
\end{equation}
To proceed, we make use of eq. (\ref{Eq/Pg10/3:dyn_radion_aDs4}). The
quantity $h'(0)$ entering that equation can be found by using  eq.
(\ref{Eqn/Pg7/6:dyn_radion_aDs4}) which takes the following form in the
presence of matter:
\[
-\sg\left(\frac{h'}{\sg} \right)'=\frac{1}{M^{d}}\left(T_{\xi \xi
}+\frac{T}{dk^{2}\sg^{2}} \right)\,,
\]
where $T\equiv g^{AB}T_{AB}$. Integrating this equation with the
boundary condition (\ref{Eq/Pg10/1:dyn_radion_adS4_2_branes}) and plugging
the result into eq. (\ref{Eq/Pg10/3:dyn_radion_aDs4}) and then into eq.
(\ref{Eq/Pg24/2A:dyn_radion_adS4_2_branes}), one finally arrives at the
desired equation of motion for the canonically normalized radion
(\ref{Eq/Pg21/1B:dyn_radion_adS4_2_branes}),
\begin{eqnarray}
&&\Box \phi -\frac{(d+1)(d+3)}{4\tau ^{2}}\phi = \nonumber\\
&&\sqrt{\frac{{|\mathcal{ P}|H^{2}}}{4M^{2d}|H\tau |^{d-1}}}
\left[\frac{1}{d\sigma }\left(\mathrm{c}_{+}k_{+}\mathcal{
T}-\frac{\triangle T_{\xi \xi }}{\tau ^{2}} \right) +\frac{1}{\tau
^{2}}\int \limits_{-\xi _{h}}^{0}\frac{d\xi }{\sg_{-}k_{-}}\left(T_{\xi
\xi }^{-}+\frac{T^{-}}{dk_{-}^{2}\sg_{-}^{2}} \right)\right] \,.
\label{Eqn/Pg25/1:dyn_radion_adS4_2_branes}
\end{eqnarray}
This reiterates that the radion has unsuppressed coupling to matter
residing on the visible brane. Equation
(\ref{Eqn/Pg25/1:dyn_radion_adS4_2_branes}) shows also that the radion
does not interact with matter residing in the ``+'' region outside the
visible brane. The latter property is consistent with the fact that the
non-trivial part of the radion wave function is concentrated on and
between the branes, see the discussion after
eq.~(\ref{Eq/Pg16/1A:dyn_radion_adS4_2_branes}).

\subsection{Graviton effective action}
\label{Subsec/Graviton effective action}
In the same way one gets the
graviton effective action
\[
S_{TT}=\frac{M_{\mathrm{Pl}}^{d-1}}{4}\int
\limits_{}^{}dx^{d+1}\frac{(\partial _{\mu }h_{(d/2)ij}^{TT})^{2}}{|H\tau
|^{d-1}}\,,
\]
with $(d+1)$-dimensional Planck mass
\begin{equation}
M_{\mathrm{Pl}}^{d-1}=2M^{d}H^{d-1}\int \limits_{-\xi _{h}}^{\infty }
\frac{d\xi}{k^{d}\sg^{d}}\,,
\label{Eq/Pg23/2:dyn_radion_adS4_2_branes}
\end{equation}
where we have set $\beta _{\frac{d}{2}}^{2}=1$ which is appropriate from
the viewpoint of a $(d+1)$-dimensional observer localized on the visible
brane (see the discussion in Ref.~\cite{Randall:1999ee}). In the static
limit (\ref{Eq/Pg20/1D:dyn_radion_adS4_2_branes}), the $(d+1)$-dimensional
Planck mass is
\[
M_{\mathrm{Pl}}^{d-1}=\frac{1}{r_{h}^{d-1}k_{-}(d-1)} \left(1-
r_{h}^{d-1}\frac{\triangle k}{k_{+}} \right)\,.
\]
This agrees with Ref.~\cite{Lykken:1999nb}.

\subsection{Limit of single visible brane}
\label{Subsec/Pg26/1:dyn_radion_adS4_2_branes/Transition to single brane}
\subsubsection{$k_{-}>0$} In the case $k_{-}>0$, the adS boundary is
located at $-\xi _{-}<-\xi _{h}<0$ and the hidden brane can be pushed to
it, $\xi _{h}\to \xi _{-}$. In this limit one has
\[
\mathcal{ P}=4d(d+1)\frac{M^{d}}{H}\left(
(d+1)\frac{k_{+}\mathrm{c_{+}}}{k_{-}\mathrm{c}_{-}}\frac{H}{\sigma }
+\left(\frac{\mathrm{s_{-}}}{\mathrm{c_{-}}}
\right)^{3}F\left(1,\frac{3}{2};\frac{d+3}{2};
\frac{\mathrm{s}_{-}^2}{\mathrm{c}_{-}^2} \right)\right)^{-1}\, .
\]
This is finite and, therefore, the radion does not decouple from the
physical spectrum. The radion-matter coupling (\ref
{Eqn/Pg25/1:dyn_radion_adS4_2_branes}) is finite as well. On the other
hand, the integral (\ref{Eq/Pg23/2:dyn_radion_adS4_2_branes}) that yields
the effective Planck mass, diverges and hence the graviton does not
interact with matter and decouples.

\subsubsection{$k_{-}<0$} In the opposite case $k_{-}<0$ the adS boundary
is absent ($\xi _{-}<0$) and the single brane limit corresponds to $\xi
_{h}\to \infty $.  In that case only the last term in the expression for
the radion wave function in the ``$-$'' domain
(\ref{Eq/Pg15/1:dyn_radion_adS4_2_branes}) survives. The radion becomes
pure gauge, and hence unphysical, in both domains.
One can also  see  that in the limit $\xi _{h}\to \infty $, $\mathcal{ P}$
vanishes. So, the radion does not couple to matter, as it should be.

On the contrary, the effective Planck mass
(\ref{Eq/Pg23/2:dyn_radion_adS4_2_branes}) is finite and graviton is the
only light physical degree of freedom.

\section{Conclusion}
\label{Section/Pg32/1:dr/Conclusion}

To conclude, in this paper we have performed the analysis of the
linearized metric perturbations in the dynamical Lykken-Randall type
model. We have derived  equations of motion for the scalar, vector and
tensor modes and have shown that, in general, the radion and graviton are
the only light modes. However, in the single brane regime, depending on
the behaviour of the warp factor in the ``$-$'' region, graviton or radion
decouples from the physical spectrum: if the warp factor grows outward the
visible brane ($k_{-}>0$) and there is the adS boundary, only the radion
is present in the physical spectrum while the graviton decouples, and vise
versa in the opposite case. We have also shown that  if the visible brane
has  negative tension, the radion is a ghost. Although these features of
the metric perturbations are interesting by themselves, we think our main
result is the radion equation of motion. This equation leads to the red
power spectrum, as one could have anticiated from the holographic picture.
This means that the potentially observable features of the
(pseudo)conformal Universe ~\cite{Libanov:2015iwa} hold also for the de
Sitter brane moving in the adS background.

\section*{Acknowledgements} The authors are indebted to E.~Nugaev and
S.~Sibiryakov for useful comments and discussions.  This work was
supported by the Russian Science Foundation grant~14-12-01430.

\appendix
\section{Spectrum of the operator (\ref{Eq/Pg16/3:dyn_radion_adS4})}
\label{Appendix Spectrum}

Let us find the spectrum of eigenvalues $\kappa^2$ in
eq.~(\ref{Eq/Pg16/3:dyn_radion_adS4}). The eigenfunctions $\beta_\kappa$
and their first derivatives must be continuous across the visible brane
and obey $\beta_\kappa^\prime (-\xi_h) = 0$ at the hidden brane.

We multiply eq.~(\ref{Eq/Pg16/3:dyn_radion_adS4}) by $\beta _{\kappa
}^{*}$, integrate the result with the measure $1/(k\sg)^{d}$, and, taking
into account the boundary conditions, obtain
\[
\int \limits_{-\xi _{h}}^{\infty }\frac{d\xi }{(k\sg)^{d}}
|\beta_{\kappa }'|^{2}=\int \limits_{-\xi _{h}}^{\infty } \frac{d\xi
}{(k\sg)^{d}} \left(\frac{d^{2}}{4} -\kappa ^{2}\right)|\beta _{\kappa
}|^{2}\,,
\]
which shows that $\kappa^2\leq d^{2}/4$. As we will see, the spectrum
is continuous at $\kappa ^{2}\leq 0$. At $\kappa =d/2$, there is the
constant mode (\ref{Eq/Pg13/1:dyn_radion_adS4_2_branes}). We will argue
that the latter mode disappears for $k_{-}>0$ and $\xi _{h}\to \xi _{-}$,
that is, when the hidden brane is pushed to the adS boundary.

There may exist  solutions with
\[
d^{2}/4>\kappa ^{2}>0\,.
\]
Our main purpose here  is to demonstrate that, in fact, there are no
such solutions. To this end we introduce the wave function
\[
\tilde{\beta }_{\kappa }(\xi )=\beta_{\kappa } (\xi ) (k\sg)^{-d/2}
\]
and cast eq. (\ref{Eq/Pg16/3:dyn_radion_adS4})  into the form of the
Schr\"odinger equation
\begin{equation}
-\partial _{\xi }^{2}\tilde{\beta }_{\kappa
}+\left(\frac{d(d+2)}{4\sg^{2}}-\frac{d\sigma }{2H}\delta (\xi )
\right)\tilde{\beta }_{\kappa }=-\kappa ^{2}\tilde{\beta }_{\kappa }\,,
\label{Eq/Pg19/1:dyn_radion_adS4}
\end{equation}
where the appearance of $\delta (\xi )$ is due to the continuous matching
conditions for $\beta_{\kappa } $ on the visible brane which translates to
the following conditions for $\tilde{\beta }_{\kappa }$,
\begin{equation}
\triangle\tilde{\beta }_{\kappa }=0\,,\ \ \ \ \triangle\tilde{\beta
}'_{\kappa }=-\frac{d\sigma }{2H}\tilde{\beta }_{\kappa }\,.
\label{Eq/Pg19/2:dyn_radion_adS4}
\end{equation}
The boundary condition on the hidden brane
(\ref{Eq/Pg10/1:dyn_radion_adS4_2_branes}) takes the following form,
\begin{equation}
\tilde{\beta }'_{\kappa
}+\frac{d\mathrm{c}_{h}}{2\mathrm{s}_{h}}\tilde{\beta }_{\kappa }=0\,,
\label{Eq/Pg19/1:dyn_radion_adS4_2_branes}
\end{equation}
and $\tilde{\beta }_{\kappa }$ should be normalizable with unit measure:
\begin{equation}
\int \limits_{-\xi _{h}}^{\infty }d\xi \tilde{\beta }_{\kappa
'}^{*}\tilde{\beta }_{\kappa }=\delta _{\kappa ',\kappa }\,,
\label{Eq/Pg20/1:dyn_radion_adS4_2_branes}
\end{equation}
where $\delta _{\kappa ',\kappa }=\delta (\kappa '-\kappa )$ for the modes
belonging to the continuous part of the spectrum.

To warm up, let us demonstrate that at $\kappa ^{2}< 0$ the spectrum is
continuous. In general, in the ``$-$'' region, there  always exist two
linear independent solutions to eq.~(\ref{Eq/Pg19/1:dyn_radion_adS4}), and
hence one can construct unique solution (up to an overall constant)
$\tilde{\beta }_{\kappa }^{<}$ to eq.~(\ref{Eq/Pg19/1:dyn_radion_adS4})
satisfying the boundary condition
 (\ref{Eq/Pg19/1:dyn_radion_adS4_2_branes}) on the hidden brane. At large
$\xi >0$, the potential term in eq. (\ref{Eq/Pg19/1:dyn_radion_adS4}) can
be neglected and at $\kappa ^{2}<0$ there are two oscillating solutions
 $\sim \mathrm{e}^{\pm i|\kappa|\xi  }$. A linear combination of them can
be chosen to satisfy eq. (\ref{Eq/Pg19/2:dyn_radion_adS4}) (at any $\kappa
^{2}<0$) and to match $\tilde{\beta }_{\kappa }^{<}$. Thus, the spectrum
is indeed continuous at $\kappa^2 < 0$. This argument does not apply to
the special case $\kappa=0$ when the asymptotic behaviour of the two
solutions at $\xi \to \infty $  is $\mathrm{const}\neq 0$ and $\xi $,
since only the first one is suitable. In any case, if $\tilde{\beta }_{0}$
exists then it belongs to the continuous part of the spectrum.

To see that the boundary value problem (\ref{Eq/Pg19/1:dyn_radion_adS4})
-- (\ref{Eq/Pg20/1:dyn_radion_adS4_2_branes}) has only one discrete
solution, we note that the first term in parenthesis in eq.
(\ref{Eq/Pg19/1:dyn_radion_adS4}) is always positive $V\propto
1/\sg^{2}>0$. Let us turn off this term. Then we deal with a particle in
the $\delta $-function well. It is straightforward to check that the
spectrum in that case consists of one negative discrete level and
continuous part starting from zero. Switching on  $V$ in
(\ref{Eq/Pg19/1:dyn_radion_adS4}) can only lead to a non-negative addition
to each eigenvalue. Since the continuous parts coincide in  both cases
(vanishing and nonvanishing $V$) this means that non-zero potential may
lead to the disappearance of the negative discrete level, but it cannot
lead to the appearance of the second negative discrete level. Therefore,
the boundary value problem (\ref{Eq/Pg19/1:dyn_radion_adS4}) --
(\ref{Eq/Pg20/1:dyn_radion_adS4_2_branes}) can have only one discrete
level and, indeed, it has the level with $\kappa =d/2$.

Let us consider the case of the single visible brane. In general, there
are two different cases: $k_{-}\leq 0$ ($\xi _{-}\leq 0$) and $k_{-}>0$
($\xi _{-}>0$). In the first case the boundary condition on the hidden
brane (\ref{Eq/Pg19/1:dyn_radion_adS4_2_branes}) is replaced by the
normalization condition (\ref{Eq/Pg20/1:dyn_radion_adS4_2_branes}) with
$\xi _{h}\to -\infty $. It is straightforward to see that  all above
arguments are still in force in that case. So, the spectrum consists of
the discrete level with $\kappa =d/2$ and continuous part starting from
zero.

In the case $k_{-}>0$ one replaces $\xi _{h}\to \xi _{-}$ and the boundary
condition (\ref{Eq/Pg19/1:dyn_radion_adS4_2_branes}) becomes
\[
\tilde{\beta }_{\kappa }(-\xi _{-})=0\,,
\]
that is, the wave functions vanish at the adS boundary, and the above
arguments do not work. Let us argue that there are no discrete levels in
this case.

Suppose that there exists a discrete level. The corresponding wave
function, being the wave function of the ground state, has no nodes and
can be chosen to be positive everywhere. Then, integrating eq.
(\ref{Eq/Pg19/1:dyn_radion_adS4}) and taking into account the boundary and
matching conditions, one obtains the following inequality,
\begin{equation}
\frac{d\sigma }{2H}\tilde{\beta }_{\kappa }(0)=\int \limits_{-\xi
_{-}}^{\infty }d\xi \left(\frac{d(d+2)}{4\sg^{2}}+\kappa ^{2}
\right)\tilde{\beta }_{\kappa }>\int \limits_{0}^{\infty }d\xi
\frac{d(d+2)}{4\sg^{2}_{+}}\tilde{\beta }_{\kappa }\,.
\label{Eq/Pg21/2:dyn_radion_adS4_2_branes}
\end{equation}
Let us consider two extreme cases: a) $\xi _{+}\ll 1$ and b) $\xi _{+}\gg
1$. The first case (see (\ref{Eq/Pg10/2:dyn_radion_aDs4})) corresponds to
slowly expanding brane, $H/k_{+}\ll 1$, and hence $\xi _{+}\simeq
H/k_{+}$. In that case the integral in the right hand side of
eq.~(\ref{Eq/Pg21/2:dyn_radion_adS4_2_branes}) is saturated near the
origin and is proportional to $1/\xi _{+}$:
\[
\frac{d\sigma }{2H}\tilde{\beta }_{\kappa
}(0)>\frac{d(d+2)}{4}\frac{1}{\xi _{+}}\tilde{\beta }_{\kappa
}(0)=\frac{d(d+2)}{4}\frac{k_{+}}{H}\tilde{\beta }_{\kappa }(0)\,,
\]
or
\[
1>\frac{(d+2)}{2}\frac{k_{+}}{\sigma }\, .
\]
This contradicts the relation $k_{+}>\sigma $ which follows from
(\ref{Eq/Pg9/5:dyn_radion_aDs4}). Hence, there is no discrete level in
that case.

The opposite case $\xi _{-}>\xi _{+}\gg 1$ corresponds to rapidly
expanding brane, $H\gg k_{\pm},\sigma $. In that case
\begin{equation}
\xi _{\pm}\simeq \log \left(\frac{2H}{k_{\pm}} \right)\,,
\label{Eq/Pg22/1:dyn_radion_adS4_2_branes}
\end{equation}
and the first term in parenthesis in eq.~(\ref{Eq/Pg19/1:dyn_radion_adS4})
can be neglected. Indeed, in the ``$+$'' region this approximation is
valid at all $\xi $, while in the ``$-$'' region the approximation may
only decrease the value of $\kappa^2$. Then, solving eqs.
(\ref{Eq/Pg19/1:dyn_radion_adS4}), (\ref{Eq/Pg19/2:dyn_radion_adS4})  with
vanishing potential, one finds
\begin{eqnarray}
&&\kappa =\frac{d\sigma }{4H}\left(1-\mathrm{e}^{-2\xi _{-}\kappa }
\right)\simeq \frac{d\sigma }{4H}\ll 1\,,\nonumber\\
&& \tilde{\beta }_{\kappa }(\xi >0)=\tilde{\beta }_{\kappa
}(0)\mathrm{e}^{-\kappa \xi }\,.
\label{Eq/Pg22/2:dyn_radion_adS4_2_branes}
\end{eqnarray}
By substituting (\ref{Eq/Pg22/1:dyn_radion_adS4_2_branes}),
(\ref{Eq/Pg22/2:dyn_radion_adS4_2_branes}) into
(\ref{Eq/Pg21/2:dyn_radion_adS4_2_branes}) one obtains
\[
1>\frac{(d+2)}{2}\frac{k_{+}^{2}}{H\sigma (2+\kappa
)}\simeq\frac{(d+2)}{4}\frac{k_{+}^{2}}{H\sigma}>\frac{d+2}{2}\,,
\]
where we have used (\ref{Eq/Pg10/1:dyn_radion_aDs4})  and inequality
\[
((k_{+}+k_{-})^{2}-\sigma ^{2})((k_{+}-k_{-})^{2}-\sigma ^{2})<
(k_{+}+k_{-})^{2}(k_{+}-k_{-})^{2}\leq k_{+}^{4}\,.
\]
Thus, we again come to contradiction  and the discrete level is absent.

Another way to see that the discrete level is absent is to consider what
happens with the mode $\kappa =d/2$ in the limit $\xi _{h}\to -\xi _{-}$.
In this limit the normalized mode has the form,
\begin{equation}
\tilde{\beta }_{\frac{d}{2}}(\xi ) =\sqrt{d-1}(\xi _{-}-\xi
_{h})^{\frac{d-1}{2}} \frac{k_{-}^{\frac{d}{2}}}{(k\sg)^{\frac{d}{2}}}\,,
\label{Eq/Pg22/4:dyn_radion_adS4_2_branes}
\end{equation}
where we have used the fact that the corresponding normalization integral
is saturated at $\xi \to \xi _{h}$:
\[
\int \limits_{-\xi _{h}\to -\xi _{-}}^{\infty }\frac{d\xi
}{(k\sg)^{d}}\simeq \frac{1}{(d-1)k_{-}^{d}(\xi _{-}-\xi _{h})^{d-1}}\,.
\]
As we have discussed above at any $\xi _{h}\neq \xi _{-}$ the mode
(\ref{Eq/Pg22/4:dyn_radion_adS4_2_branes}) is the only discrete mode in
the spectrum. It follows from (\ref{Eq/Pg22/4:dyn_radion_adS4_2_branes})
that at any given $\xi >\xi _{-}$ this mode tends to zero in the limit
$\xi _{h}\to \xi _{-}$ and, therefore, does not contribute to any
observable in the whole space except for an infinitesimal region near the
adS boundary.

To summarize, we have seen that the spectrum of the operator
(\ref{Eq/Pg16/3:dyn_radion_adS4}) defined on the class of continuous
functions in the case of  two branes as well as in the case of single
brane and $k_{-}\leq 0$  consits of one discrete level with $\kappa =d/2$
($\nu =1-d$) and continuous part starting from  $\kappa=0$, $\nu
=(d-2)^{2}/4$. In the case of single brane and $k_{-}>0$ the discrete
level is absent, the spectrum is continuous and starts from $\kappa=0$,
$\nu =(d-2)^{2}/4$.

\end{document}